\documentclass[]{aa} 
\usepackage{txfonts}
\usepackage{natbib}
\usepackage{graphicx, xspace}
\usepackage{placeins}
\usepackage{array}
\usepackage[switch]{lineno}
\usepackage[draft]{hyperref}  
\usepackage{color}
\usepackage{gensymb}
\usepackage{amsmath}
\usepackage{longtable}

\newcommand{\Lagr}{\mathcal{L}}
\newcommand{\Magr}{\mathcal{M}}
\newcommand\omicron{o}
\def\d{\,d$^{-1}$\xspace}
\def\kms{km\,s$^{-1}$\xspace}
\def\OmiLup{$\omicron$\,Lup\xspace}

\makeatletter

\newcommand{\Rmnum}[1]{\expandafter\@slowromancap\romannumeral #1@}
\makeatother

\begin{document}


\title{Magnetic characterization and variability study of the magnetic SPB star
  $\omicron$\,Lup\thanks{This work was based on data gathered with HARPS
    installed on the 3.6-m ESO telescope (ESO Large Programme 187.D-0917 and ESO
    Normal Programme 097.D.0156) at La Silla, Chile and on observations obtained 
    at the Canada-France-Hawaii Telescope (CFHT) which is operated by the National 
    Research Council of Canada, the Institut National des Sciences de l'Univers of 
    the Centre National de la Recherche Scientifique of France, and the University of 
    Hawaii..}$^{,}$\thanks{Mode
    identification results obtained with the software package FAMIAS developed
    in the framework of the FP6 European Coordination Action HELAS
    (http://www.helas-eu.org).}$^{,}$\thanks{Based on data collected by the
    BRITE Constellation satellite mission, designed, built, launched, operated
    and supported by the Austrian Research Promotion Agency (FFG), the
    University of Vienna, the Technical University of Graz, the Canadian Space
    Agency (CSA), the University of Toronto Institute for Aerospace Studies
    (UTIAS), the Foundation for Polish Science \& Technology (FNiTP MNiSW), and
    National Science Centre (NCN).}}

\authorrunning{B.~Buysschaert}
\titlerunning{Magnetic characterization and variability of $\omicron$\,Lup}

\author{B.\,Buysschaert\inst{1,2}, 
C.\,Neiner\inst{1},
A.\,J.\,Martin\inst{1},
M.\,E.\,Oksala\inst{3,1},
C.\,Aerts\inst{2,4},
A.\,Tkachenko\inst{2},
E.\,Alecian\inst{5}, \and \\
the MiMeS Collaboration
}
\offprints{bram.buysschaert@obspm.fr} 
\mail{bram.buysschaert@obspm.fr} 

\institute{ LESIA, Observatoire de Paris, PSL Research University, CNRS, Sorbonne Universit\'es, UPMC Univ. Paris 06, Univ. Paris Diderot, Sorbonne Paris Cit\'e, 5 place Jules Janssen, F-92195 Meudon, France 
  \and Instituut voor Sterrenkunde, KU Leuven, Celestijnenlaan 200D, 3001 Leuven, Belgium 
  \and Department of Physics, California Lutheran University, 60 West Olsen Road \# 3700, Thousand Oaks, CA, 91360, USA 
  \and Dept. of Astrophysics, IMAPP, Radboud University Nijmegen, 6500 GL, Nijmegen, The Netherlands 
  \and Universit\'e Grenoble Alpes, CNRS, IPAG, F-38000 Grenoble, France
} \abstract {Thanks to large dedicated surveys, large-scale magnetic fields have been detected for about 10\,\% of early-type stars.  We aim to precisely characterize the large-scale magnetic field of the magnetic component of the wide binary \OmiLup, by using high-resolution ESPaDOnS and HARPSpol spectropolarimetry to analyse the variability of the measured longitudinal magnetic field.  In addition, we investigate the periodic variability using space-based photometry collected with the BRITE-Constellation by means of iterative prewhitening. The rotational variability of the longitudinal magnetic field indicates a rotation period $P_{\mathrm{rot}}=2.95333(2)$\,d and that the large-scale magnetic field is dipolar, but with a significant quadrupolar contribution.  Strong differences in the strength of the measured magnetic field occur for various chemical elements as well as rotational modulation for Fe and Si absorption lines, suggesting a inhomogeneous surface distribution of chemical elements.  Estimates of the geometry of the large-scale magnetic field indicate $i=27\pm 10\,^{\circ}$, $\beta = 74^{+7}_{-9}\,^{\circ}$, and a polar field strength of at least 5.25\,kG.  The BRITE photometry reveals the rotation frequency and several of its harmonics, as well as two gravity mode pulsation frequencies.  The high-amplitude g-mode pulsation at $f=1.1057$\,\d dominates the line-profile variability of the majority of the spectroscopic absorption lines.  We do not find direct observational evidence of the secondary in the spectroscopy.  Therefore, we attribute the pulsations and the large-scale magnetic field to the B5IV primary of the \OmiLup system, but we discuss the implications should the secondary contribute to or cause the observed variability.}


\maketitle

\section{Introduction}
\label{sec:Introduction}
\subsection{Magnetic and pulsating early-type stars}
\label{sec:intro_general}
Large-scale magnetic fields are detected at the stellar surface of about 10\,\% of the studied early-type stars by measuring their Zeeman signature in high-resolution spectropolarimetry (e.g., MiMeS, \citet{2016MNRAS.456....2W}; the BOB campaign, \citet{2015IAUS..307..342M}; and the BRITE spectropolarimetric survey, \citet{2016arXiv161103285N}).  These large-scale magnetic fields appear to be stable over a time scale of decades, have a rather simple geometry (most often a magnetic dipole), and have a polar strength ranging from about 100\,G to several tens of kG.  Because the fields remain stable and their properties do not depend on any observed stellar parameters, we expect that these large-scale magnetic fields were produced during earlier stages of the star's life, relaxing into the observed configuration \citep[e.g.,][]{1999stma.book.....M, 2015IAUS..305...61N}.  In addition, a dynamo magnetic field is likely to occur in the deep interior of early-type stars, produced by the convective motions of ionized matter in the convective core \citep{1989MNRAS.236..629M}.  However, no direct evidence of such a magnetic dynamo has ever been observed at the stellar surface, nor is it expected, since the Ohmic diffusion time scale from the core to the surface is longer than the stellar lifetime.

The large-scale magnetic fields detected at the surface of early-type stars have implications on the properties of the circumstellar environment, the stellar surface, and the interior, altering the star's evolution:
\begin{itemize}
\item The ionized wind material follows the magnetic field lines, and can create (quasi-)stable structures in the circumstellar environment or magnetosphere.  The precise properties of the magnetospheric material depend on the stellar and magnetic properties \citep[e.g.,][]{2002ApJ...576..413U, 2005MNRAS.357..251T}.  In general, magnetospheres are subdivided into centrifugal magnetospheres, where material remains trapped by the magnetic field and supported against gravity by rapid rotation, and dynamical magnetospheres.  The latter has a region of enhanced density in the circumstellar environment that continuously accumulates new wind material and loses matter to accretion by the star.

\item At the stellar surface, the large-scale magnetic field can affect the stratification and diffusion of certain chemical species at the surface, which can cause surface abundance inhomogeneities of certain chemical elements, and a peculiar global photospheric abundance composition. This would lead to rotational modulation of line profiles and photometric variability. Stars for which such peculiarities are observed are denoted by the Ap/ Bp spectral classification.

\item The structure and evolution of the deep stellar interior is anticipated to be altered by the large-scale magnetic field, due to the competition of the Lorentz force with the pressure force and gravity.  This leads to a uniformly rotating radiative envelope \citep[e.g.,][]{1937MNRAS..97..458F, 1992MNRAS.257..593M, 1999A+A...349..189S, 2005A+A...440..653M, 2011IAUS..272...14Z}, altering the depth over which material overshoots the convective core boundary into the radiative layer \citep[e.g.,][]{1981ApJ...245..286P, 2004ApJ...601..512B}.  Currently, this effect has only been determined for two stars with a technique referred to as magneto-asteroseismology, which combines the analysis of the star's pulsations with that of its magnetic properties, and then performing forward seismic modelling.  This was done for the magnetic $\beta$\,Cep pulsator V\,2052\,Oph \citep{2012A+A...537A.148N, 2012MNRAS.424.2380H, 2012MNRAS.427..483B} and the magnetic g-mode pulsator HD\,43317 \citep{2017A+A...605A.104B, 2018arXiv180500802B}.
\end{itemize}

Whenever present, the properties of the stellar pulsations depend on the strength, geometry and orientation of the large-scale magnetic field \citep[e.g.,][]{1982MNRAS.201..619B, 1984MmSAI..55..215G, 1985ApJ...296L..27D, 1990MNRAS.242...25G, 1992ApJ...395..307G, 1993PASJ...45..617S, 1995PASJ...47..219T, 1996ApJ...458..338D, 2000A+A...356..218B, 2005A+A...444L..29H, 2011A+A...526A..65M, 2017MNRAS.466.2181L}.  For stars with a spectral type from O9 to B2, $\beta$\,Cep-type pulsations are expected.  These are low-order pressure modes with periods of the order of several hours.  For slightly less massive stars (spectral types B2 to B9) Slowly Pulsating B-type (SPB) oscillations are predicted.  These are low-degree, high-order gravity modes with a period of the order of a few days.  Moreover, their pulsation modes show a regular pattern in the period domain.  Both the $\beta$\,Cep and SPB pulsations are driven by the $\kappa$-mechanism, related to the temperature dependent opacity of iron-like elements.  In addition gravito-intertial modes, which are excited by the motions of the convective core and have the Coriolis force and buoyancy as restoring forces, are anticipated for early-type stars with periods longer than the stellar rotation period \citep[e.g.,][]{2014A+A...565A..47M}.  Stellar pulsations remain the sole way to probe the interior of a single star, with the parameters of the pulsation modes dependent on the conditions inside the star.

Variability due to gravity waves \citep[e.g.,][]{1981ApJ...245..286P, 2013ApJ...772...21R} was recently also found in photospheric and wind lines of the O9Iab star HD\,188209 \citep{2017A+A...602A..32A}, the B1Ia star HD\,2905 \citep{2018A+A...612A..40S}, and the B1Iab star $\rho$\,Leo \citep{2018MNRAS.476.1234A}, as well as in the close binary V380\,Cyg \citep{2014MNRAS.438.3093T}. The presence of gravity waves seems to be a common property of hot massive stars that have evolved beyond half of the core-hydrogen burning stage, irrespective of binarity or a  magnetic field.
 
\begin{figure*}[t]
		\centering
			\includegraphics[width=\textwidth, height = 0.33\textheight]{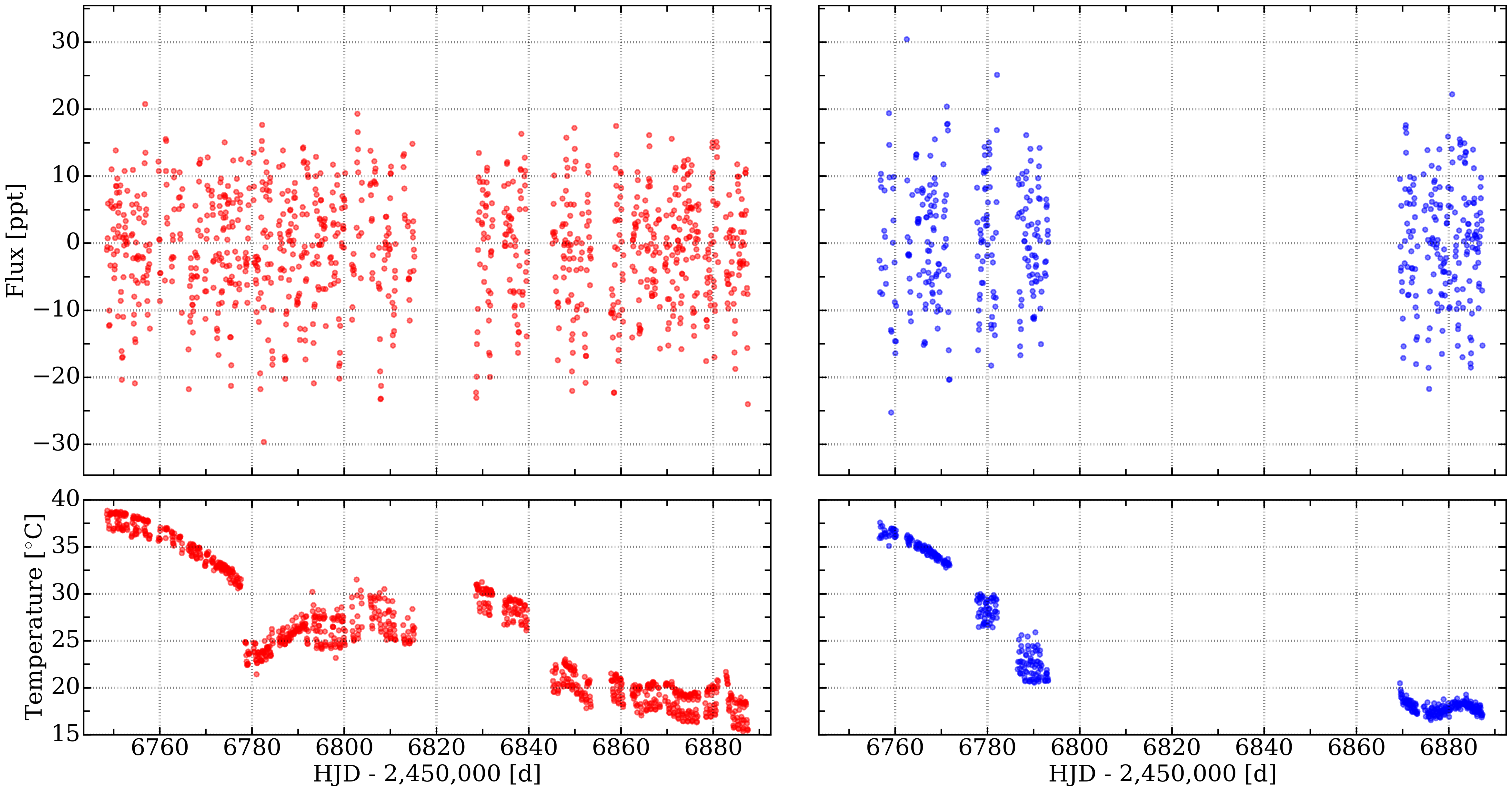}%
			\caption{\textit{Top}: Satellite-orbit averaged and reduced BRITE light curves for \OmiLup, where the UBr photometry is indicated in red (\textit{left}) and the BAb photometry in blue (\textit{right}).  Photometric variability is indicated in parts-per-thousand (ppt).  \textit{Bottom}: Corresponding satellite-orbit averaged temporal variability of the on-board CCD temperature, showing different and discontinuous behaviour.}
			\label{fig:BRITE_lightcurves}
\end{figure*}

\subsection{omicron Lupi}
\label{sec:intro_omilup}
\OmiLup (HD\,130807, HR\,5528, HIP\,72683, B5IV, $V=4.3$\,mag) is an early-type star and a member of the Sco-Cen association, following its Hipparcos parallax \citep{2011MNRAS.416.3108R}.  This region is a site of recent massive star formation at a distance of 118 -- 145\,pc, with the exact value depending on the sub-group of the association.  Isochrone fitting to the Hertzsprung-Russell diagram indicates that the star formation occurred some 5 -- 20\,Myr ago.

Using interferometry, \citet{1951CiUO..112...94F} detected a secondary component for \OmiLup, at an angular separation of 0.115\,arcsec.  The most recent interferometric measurement indicated that the components have an angular separation of 0.043\,arcsec, with a contrast ratio of $0.28 \pm 0.06$\,mag \citep{2013MNRAS.436.1694R}.  From the distance to the Sco-Cen association, the authors deduced that the components are 5.33\,au apart with a mass ratio of 0.91.  Moreover, the distance to the Sco-Cen association implies that the largest measured angular separation is above 17\,au, such that the orbital period of the binary must be longer than 20\,years \citep{2011A+A...536L...6A}.

Within the scope of the MiMeS survey, HARPSpol observations were collected for \OmiLup.  \citet{2011A+A...536L...6A} concluded that \OmiLup hosts a large-scale magnetic field, with variability of the measured longitudinal magnetic field indicating a rotation period between one and six days.  This agrees well with the small value of the projected rotation velocity, $v\sin i = 27 \pm 3$\,\kms \citep[determined by][]{2005ESASP.560..571G}.  Moreover, \citet{2011A+A...536L...6A} determined $T_{\mathrm{eff}} = 18000$\,K and $\log g = 4.25$\,dex for \OmiLup from a comparison with synthetic spectra using \textsc{tlusty} non-local thermal equilibrium atmosphere models and the \textsc{synspec} code \citep{Lanz2007, Hubeny2011}.  The authors also noted weaker \ion{He}{I} lines and stronger \ion{Si}{II} than expected from the solar abundances.  Hence, the surface abundance of certain chemical elements seems to be peculiar.  Lastly, Si, N, and Fe exhibited line-profile variations (LPVs) on a time scale of about one day. \citet{2011A+A...536L...6A} proposed surface abundance inhomogeneities as the cause of these LPVs.

\OmiLup has recently been observed by the BRITE-Constellation of nano-satellites to monitor its photometric variability.  This space-based photometry could aid in the determination of the rotation period of \OmiLup by observing the rotational modulation caused by surface abundance inhomogeneities due to the large-scale magnetic field.  Moreover, it might permit us to determine the precise value and the physical process causing the variability with a period of about one day that was noted by \citet{2011A+A...536L...6A} as LPVs.  Additional ground-based, high-resolution, optical spectropolarimetric data were collected to characterize the magnetic field of \OmiLup more precisely.

We introduce the various observational data sets in Sect.\,\ref{sec:observations}, and indicate how these were prepared and corrected for instrumental effects when needed.  In Sect.\,\ref{sec:st_params}, we estimate the stellar parameters of \OmiLup by fitting synthetic spectra to the observations and we search for evidence of the secondary component in the spectroscopy.  The periodic photometric variability is investigated in Sect.\,\ref{sec:photometry}, while Sect.\,\ref{sec:magnetic} covers the analysis of the large-scale magnetic field.  The sub-exposures of the spectropolarimetric sequences are employed to detect and characterize the LPVs in Sect.\,\ref{sec:LPV}.  We end this work by discussing the obtained results in Sect.\,\ref{sec:discussion} and by drawing conclusions and providing a summary in Sect.\,\ref{sec:conclusions}.

\section{Observations}
\label{sec:observations}
\subsection{BRITE photometry}
\label{sec:obs_BRITE}
\OmiLup was observed by three nano-satellites of the BRIght Target Explorer (BRITE)-Constellation \citep{2014PASP..126..573W} during the Centaurus\,I campaign.  The BAb (BRITE Austria blue) nano-satellite monitored \OmiLup from 9\,April\,2014 until 18\,August 2014, with a large time gap (of about $76$\,days) in the middle of the campaign, the UBr (UniBRITE red) nano-satellite performed continuous observations from 31\,March\,2014 until 27\,August\,2014, and the BTr (BRITE Toronto red) nano-satellite had a short campaign from 27\,June\,2014 until 3\,July\,2014.  Light curves were constructed by the BRITE-team from the raw CCD images using circular apertures \citep{2016PASP..128l5001P, 2017A+A...605A..26P}.  These raw light curves were corrected for the intrapixel sensitivity and additional metadata were added, such as aperture centroid position and on-board CCD temperature.  We retrieved these publicly available Data Reduction version 2 (DR2) data from the BRITE data archive\footnote{\url{https://brite.camk.edu.pl/pub/index.html}}.

The extracted BRITE photometry was further corrected by accounting for known instrumental trends using our in-house tools \citep[][see its appendix for explicit details]{2017A+A...602A..91B}.  Here, we provide a short summary of the applied procedure.  As a first step, we converted the timing of the observations to mid-exposure times.  Next, we subdivided the light curves according to the temporal variability of the on-board CCD temperature, $T_{\mathrm{CCD}}$, because strong discontinuities and differences in its variability were noted (see bottom panels of Fig.\,\ref{fig:BRITE_lightcurves}). For each of these data subsets, we performed an outlier rejection using the aperture centroid positions $x_{\mathrm{c}}$ and $y_{\mathrm{c}}$, the on-board temperature $T_{\mathrm{CCD}}$, the observed flux, and the number of datapoints per nano-satellite orbit.  Once all spurious data were removed, we recombined the datasets to convert the photometric variability to parts-per-thousand (ppt). The data were then again subdivided into the same subsets to correct for the fluctuating shape of the point-spread-function caused by the varying on-board temperature.  The next correction step was a classical decorrelation between the corrected flux and the other metadata (including the nano-satellite orbital phase) whenever the correlation was sufficiently strong.  This detrending procedure was performed for the complete UBr dataset and for the two BAb observing sub-campaigns (before and after the large time gap).  We could not correct the BTr data, since the very short 6\,days time span leads to uncertain instrumental correction. Thus, we did not use this BTr photometry in this work.  Finally, we applied a local linear regression filter to the corrected BRITE photometry, detrending and suppressing any remaining (instrumental) trend with a period longer than ${\sim} 10$\,days.  The last part consisted of determining satellite-orbit averaged measurements.

\begin{table}[t]
\caption{Diagnostics related to the two BRITE light curves of \OmiLup.}
\centering
\tabcolsep=6pt
\begin{tabular}{p{2.5cm}lcc}
\hline
\hline
					&		& UBr		& BAb	\\
\hline
$\mathrm{rms_{raw}}$ 		&[ppt]	& 4.57		& 3.50	\\
$\mathrm{rms_{corr}}$ 		&[ppt]	& 2.44		& 2.33	\\
length 						&[d]		& 139.0		& 130.7	\\
time gap			&[d]		& 13.4		& 76.3	\\
$D_{\mathrm{sat, raw}}$ 		&[\%]	& 70.5		& 26.3	\\
$D_{\mathrm{sat, corr}}$ 	&[\%]	& 53.1		& 24.0	\\
$D_{\mathrm{orb, raw}}$ 		&[\%]	& 17.4		& 12.8	\\
$D_{\mathrm{orb, corr}}$ 	&[\%]	& 15.6		& 10.3	\\
$N_{\mathrm{orb, corr}}$		&		& 1058		& 450	\\
\hline
\end{tabular}
\label{tab:BRITE_diag}
\tablefoot{For each light curve, we provide the rms scatter of the flux before and after correction, the length of the light curve, the length of the largest time gap, and the $D_{\mathrm{sat}}$ and $D_{\mathrm{orb}}$ duty cycles before and after correction.  The number of successful satellite orbits with observations is indicated as well.}
\end{table}

The final corrected, detrended, and satellite-orbit averaged photometry is given in Fig.\,\ref{fig:BRITE_lightcurves}.  To assess the quality of the reduced BRITE photometry, we also provide the values for some diagnostic parameters in Table\,\ref{tab:BRITE_diag}.  The first of these parameters is the root mean square (rms) of the flux, given as
\begin{equation}
\mathrm{rms} = \sqrt{\frac{1}{N} \sum\limits_{i}^{N}{\frac{\sigma_i^2}{k_i}}} \,\mathrm{,}
\label{eq:rms_flux}
\end{equation}
\noindent where $\sigma_i$ and $k_i$ are the standard deviation of the flux and the number of observations within orbital passage $i$, respectively, and $N$ is the total number of orbital passages for a given BRITE dataset.  Also listed are the median duty cycle per satellite orbit $D_{\mathrm{orb}}$ and the fraction of successful satellite orbits $D_{\mathrm{sat}}$.  The former indicates the portion of the satellite orbit used for observations, while the latter parametrizes the amount of successful satellite orbits over the total time span of the light curve.

The two final BRITE light curves (UBr and BAb) formed the basis of the photometric analysis of the periodic variability of \OmiLup discussed in Sect.\,\ref{sec:photometry}.

\subsection{Spectropolarimetry}
\label{sec:obs_specpol}
The HARPSpol polarimeter \citep{2011Msngr.143....7P} was employed in combination with the High Accuracy Radial velocity Planet Searcher (HARPS) spectrograph \citep{2003Msngr.114...20M} to measure the Zeeman signature indicating the presence of a large-scale magnetic field at the surface of \OmiLup.  This combined instrument is installed at the ESO 3.6-m telescope at La Silla Observatory (Chile) and covers the 3800--6900\,$\AA$ wavelength region with an average spectral resolution of 110\,000.  Standard settings were used for the instrument, with bias, flat-field, and ThAr calibrations taken at the beginning and end of each night.  In total, 36 spectropolarimetric sequences were obtained during three different observing runs, in May\,2011, July\,2012, and April\,2016.  The first two campaigns were part of the Magnetism In Massive Stars (MiMeS) survey \citep{2016MNRAS.456....2W}, and the third observing run was performed for the BRITE spectropolarimetric survey \citep{2016arXiv161103285N}.  Each spectropolarimetric sequence consists of four consecutive sub-exposures with a constant exposure time ranging between 207\,s and 1000\,s.  An overview of the spectropolarimetric dataset is given in Table\,\ref{tab:specpol_log}.

\begin{table*}[t]
\caption{Observing log of the spectropolarimetric sequences.}
\centering
\tabcolsep=6pt
\begin{tabular}{l|ccc|c|c|cl|c|c|cc}
\hline
\hline
ID	& HJD [d]	& $t_{\rm exp}$ [s]	& $\phi_{\rm rot}$	& complete 	&He excluded& \multicolumn{2}{c|}{\ion{He}{I}}& \ion{Fe}{II}	& \ion{Si}{II}&\multicolumn{2}{c}{Balmer}\\
	& -2450000	&		&				  				& S/N		& S/N		&S/N & Detect.					&S/N				& S/N &S/N & Detect.\\
\hline
H01	&	5704.72965	&$	4 \times 100	$&	0.754961	&	5186	&	4154	&	567	&	DD	&	2879	&	920	&	137	&	DD	\\
H02	&	5708.75948	&$	4 \times 300	$&	0.119465	&	4868	&	3959	&	595	&	ND	&	2768	&	933	&	143	&	DD	\\
H03	&	5709.73559	&$	4 \times 750	$&	0.449976	&	4875	&	3984	&	562	&	DD	&	2809	&	914	&	138	&	DD	\\
H04	&	5709.77216	&$	4 \times 750	$&	0.462360	&	4798	&	3816	&	547	&	ND	&	2752	&	897	&	143	&	DD	\\
H05	&	6123.56357	&$	4 \times 300	$&	0.572476	&	5101	&	3794	&	548	&	DD	&	2664	&	826	&	154	&	DD	\\
H06	&	6124.70439	&$	4 \times 300	$&	0.958758	&	5094	&	3862	&	627	&	DD	&	2767	&	916	&	134	&	DD	\\
H07	&	6125.45930	&$	4 \times 300	$&	0.214371	&	5300	&	4055	&	626	&	ND	&	3067	&	1012	&	140	&	ND	\\
H08	&	6125.57061	&$	4 \times 300	$&	0.252060	&	5289	&	4059	&	562	&	ND	&	3173	&	1038	&	128	&	ND	\\
H09	&	6126.56054	&$	4 \times 300	$&	0.587252	&	5130	&	3798	&	617	&	DD	&	2753	&	858	&	128	&	DD	\\
H10	&	6127.46812	&$	4 \times 300	$&	0.894560	&	5350	&	4064	&	534	&	ND	&	2943	&	955	&	147	&	DD	\\
H11	&	6129.59043	&$	4 \times 600	$&	0.613176	&	5010	&	3809	&	609	&	DD	&	2682	&	837	&	157	&	DD	\\
H12	&	6130.51806	&$	4 \times 600	$&	0.927273	&	5350	&	3997	&	581	&	DD	&	2861	&	933	&	152	&	DD	\\
E01	&	6758.06034	&$	4 \times 85	$&	0.413611	&	3214	&	2736	&	408	&	DD	&	2234	&	714	&	60	&	DD	\\
E02	&	6819.85180	&$	4 \times 85	$&	0.336251	&	3335	&	2664	&	611	&	ND	&	2383	&	676	&	60	&	DD	\\
H13	&	7481.61063	&$	4 \times 207	$&	0.408343	&	5134	&	4163	&	584	&	DD	&	3100	&	965	&	141	&	DD	\\
H14	&	7481.62175	&$	4 \times 207	$&	0.412107	&	5092	&	4041	&	582	&	DD	&	3006	&	958	&	138	&	DD	\\
H15	&	7481.63287	&$	4 \times 207	$&	0.415871	&	5251	&	4079	&	582	&	DD	&	2988	&	961	&	139	&	DD	\\
H16	&	7481.64398	&$	4 \times 207	$&	0.419634	&	5212	&	4119	&	579	&	DD	&	3039	&	960	&	158	&	DD	\\
H17	&	7481.65509	&$	4 \times 207	$&	0.423396	&	5133	&	4050	&	576	&	DD	&	2989	&	940	&	156	&	DD	\\
H18	&	7481.66620	&$	4 \times 207	$&	0.427159	&	5150	&	4035	&	576	&	DD	&	2974	&	943	&	151	&	DD	\\
H19	&	7481.67731	&$	4 \times 207	$&	0.430921	&	5099	&	4010	&	574	&	DD	&	2908	&	929	&	150	&	DD	\\
H20	&	7481.68842	&$	4 \times 207	$&	0.434684	&	4914	&	3884	&	537	&	DD	&	2857	&	922	&	154	&	DD	\\
H21	&	7481.82751	&$	4 \times 207	$&	0.481777	&	4513	&	3573	&	508	&	ND	&	2488	&	804	&	156	&	DD	\\
H22	&	7481.83862	&$	4 \times 207	$&	0.485539	&	3772	&	3138	&	589	&	ND	&	2170	&	740	&	143	&	ND	\\
H23	&	7482.65216	&$	4 \times 207	$&	0.761005	&	4810	&	3734	&	577	&	DD	&	2666	&	897	&	139	&	DD	\\
H24	&	7482.66327	&$	4 \times 207	$&	0.764767	&	4603	&	3639	&	561	&	ND	&	2585	&	883	&	159	&	MD	\\
H25$^{a}$	&	7482.67438	&$	4 \times 207	$&	0.768530	&		&		&		&		&		&		&		&		\\
H26	&	7484.69562	&$	4 \times 207	$&	0.452922	&	4728	&	3707	&	570	&	DD	&	2737	&	886	&	155	&	DD	\\
H27	&	7484.70673	&$	4 \times 207	$&	0.456684	&	4738	&	3748	&	564	&	ND	&	2772	&	890	&	160	&	DD	\\
H28	&	7484.79338	&$	4 \times 207	$&	0.486023	&	4727	&	3710	&	563	&	DD	&	2693	&	866	&	153	&	DD	\\
H29	&	7484.80449	&$	4 \times 207	$&	0.489786	&	4749	&	3726	&	546	&	DD	&	2677	&	867	&	137	&	DD	\\
H30	&	7484.88788	&$	4 \times 207	$&	0.518020	&	4720	&	3682	&	545	&	DD	&	2621	&	812	&	146	&	DD	\\
H31	&	7484.89899	&$	4 \times 207	$&	0.521782	&	4759	&	3704	&	604	&	DD	&	2627	&	808	&	156	&	DD	\\
H32	&	7485.73598	&$	4 \times 207	$&	0.805188	&	5107	&	4035	&	604	&	DD	&	2908	&	951	&	153	&	DD	\\
H33	&	7485.74709	&$	4 \times 207	$&	0.808952	&	5073	&	3971	&	601	&	DD	&	2912	&	957	&	159	&	DD	\\
H34	&	7485.83035	&$	4 \times 207	$&	0.837143	&	5101	&	3939	&	600	&	DD	&	2828	&	895	&	157	&	DD	\\
H35	&	7485.84146	&$	4 \times 207	$&	0.840905	&	5072	&	3939	&	367	&	DD	&	2830	&	909	&	161	&	DD	\\

\hline
\end{tabular}
\label{tab:specpol_log}
\tablefoot{The first letter of the ID indicates whether the spectropolarimetric sequence was taken with HARPS (H) or ESPaDOnS (E). For each sequence, the mid-exposure HJD, the exposure time, and the rotation phase, $\phi_{\rm rot}$, are indicated.  The latter was determined with $P_{\rm rot} = 2.95333$\,d and $T_0 = \mathrm{HJD}\,2455702.5$.  The provided S/N is that of the LSD Stokes\,I profile calculated with various line masks.  In addition, the magnetic detection status is provided (DD = Definite Detection, MD = Marginal Detection, and ND = Non Detection) in case not all observations resulted in a DD for the given LSD line mask.  $^{a}$ This observation was discarded because the last two sub-exposures in the sequence did not contain any signal due to bad weather.}
\end{table*}

The HARPSpol data were reduced using the \textsc{reduce} package \citep{2002A+A...385.1095P, 2011A+A...525A..97M}, and resulted in a circular spectropolarimetric observation for each sequence.  A minor update to the package enabled us to extract individual spectra for each sub-exposure of the spectropolarimetric sequence.  The spectropolarimetric observations were normalized to unity continuum using an interactive spline fitting procedure \citep[][]{2018MNRAS.475.1521M}.  This normalization method was performed per spectral order to achieve a smooth overlap between consecutive spectral orders.  For the spectroscopy employed in this work, only the spectral orders of interest, taken from the spectropolarimetric sub-exposures, were normalized with the same interactive procedure.

\OmiLup was also observed twice with the Echelle SpectroPolarimetric Device for the Observation of Stars \citep[ESPaDOnS,][]{2006ASPC..358..362D} mounted at the Canada France Hawaii Telescope (CFHT) on Mauna\,Kea in Hawaii in April and June\,2014 (PI: M. Shultz).  These spectropolarimetric sequences comprise of four consecutive sub-exposures with an exposure time of 85\,s.  They span the 3700--10500\,$\AA$ wavelength region with an average resolving power of 65\,000.  The data were reduced with the \textsc{libre-esprit} \citep{1997MNRAS.291..658D} and \textsc{upena} softwares available at CFHT.  The resulting ESPaDOnS spectropolarimetric and spectroscopic observations were normalized to unity continuum in the same manner and using the same interactive tool as for the HARPSpol data.  Details for the ESPaDOnS spectropolarimetry are provided in Table\,\ref{tab:specpol_log}.

\section{Comparison with synthetic spectra}
\label{sec:st_params}

\begin{figure*}[t]
		\centering
			\includegraphics[width=\textwidth, height = 0.33\textheight]{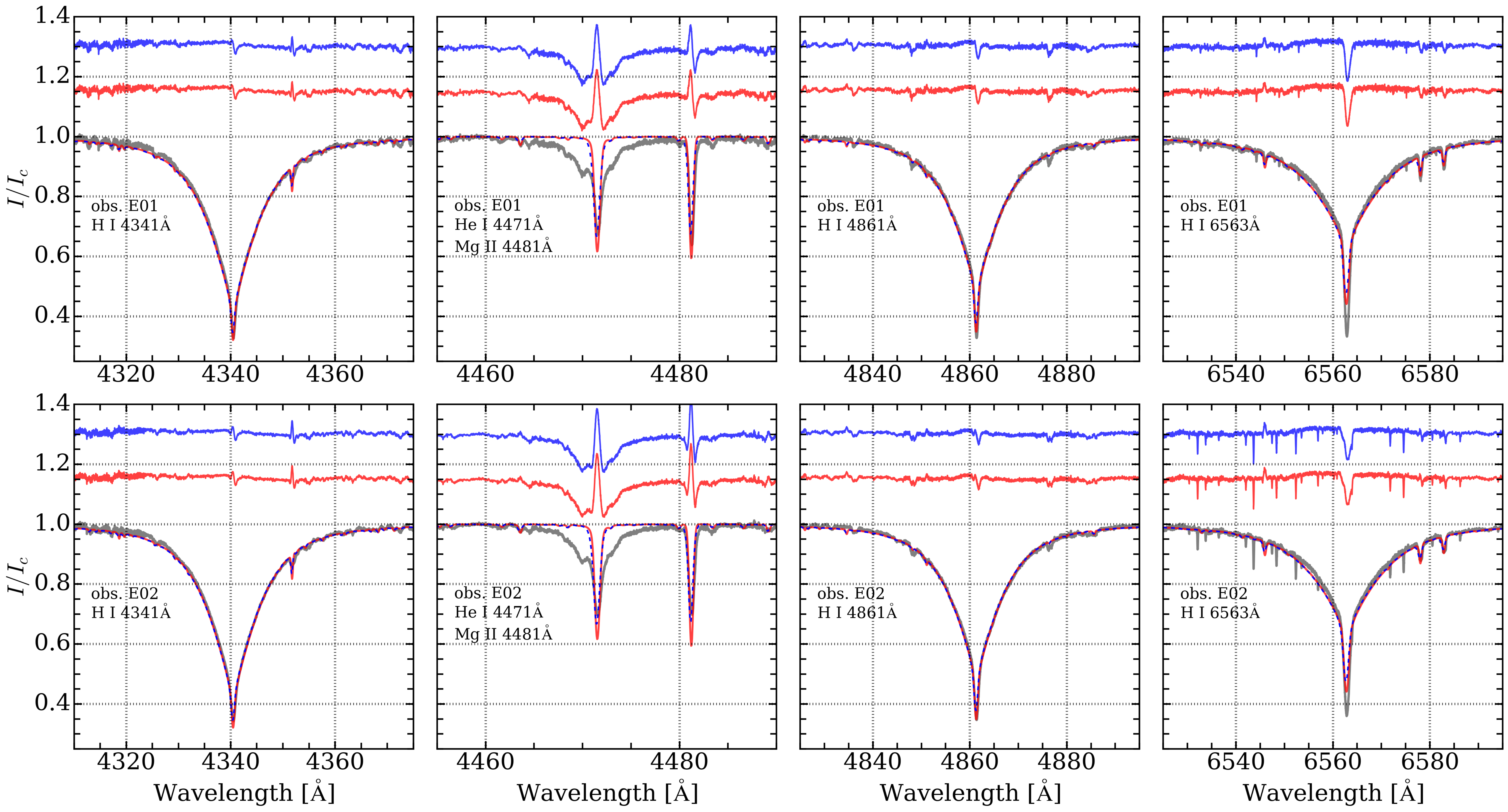}%
			\caption{Comparison between two ESPaDOnS spectra of \OmiLup, taken two months apart (shown in gray), and synthetic \textsc{atlas9}\,/\,\textsc{cossam\_simple} spectra. The red line is the synthetic spectrum for a single star with $T_{\mathrm{eff}}=15000$\,K, $\log g=3.8$\,dex, and $v \sin i = 35$\,\kms, and the blue line is the synthetic spectrum for a binary star with $T_{\mathrm{eff, 1}} = 17000$\,K, $ T_{\mathrm{eff,2 }} = 14000$\,K, $\log g_1 = 3.9$\,dex, $\log g_2 = 3.8$\,dex, $v \sin i_1 = 50$\,\kms, $v \sin i_2=25$\,\kms, $RV_1 =-5$\,\kms, $RV_2 =+5$\,\kms, and a light fraction of 50\,\%.  Residuals to the fit are indicated with the same color coding and a small offset for increased visibility.  The \textit{top} row corresponds to the observed spectrum E01 and the \textit{bottom} row to E02 (see also Table\,\ref{tab:specpol_log}).}
			\label{fig:spec_synthetic}
\end{figure*}

An accurate magnetometric analysis (Sect.\,\ref{sec:magnetic}) starts from the appropriate spectral line pattern (also referred to as the line mask), which is defined by the atmospheric characteristics of the star.  To this aim, we used a grid of synthetic spectra to model the observed Balmer lines ($\mathrm{H\alpha}$, $\mathrm{H\beta}$, $\mathrm{H\gamma}$, and $\mathrm{H\delta}$) and selected helium and metal lines, deriving a value for the effective temperature, $T_{\mathrm{eff}}$, and the surface gravity, $\log g$, of \OmiLup.  The selected lines included the \ion{He}{I}\,4471$\AA$ line and the \ion{Mg}{II}\,4481$\AA$ line, since their relative depths  are good indicators for $T_{\mathrm{eff}}$ and $\log g$.  The synthetic spectra produce a (good) first approximation of the stellar parameters, because the fainter secondary component and possible surface abundance inhomogeneities will lead to an uncertain chemical abundance analysis.

A model grid covering a range of $T_{\mathrm{eff}}$ spanning from $3500$\,K to $55000$\,K and $\log g$ from 0.00\,dex to 5.00\,dex\footnote{As $T_{\mathrm{eff}}$ increases, the range of the $\log g$ values covered reduces to values between $4.00$\,dex and $4.75$\,dex at $55000$\,K.} was calculated using \textsc{atlas9} model atmospheres by \citet[][retrieved from the Mikulski Archive for Space Telescopes (MAST)]{Bohlin2017} and \citet{Martin2017} assuming plane parallel geometry, local thermodynamic equilibrium, and an opacity distribution function for solar abundances \citep{Kurucz1993b}. Synthetic spectra were computed with \textsc{cossam\_simple} \citep{Martin2017}.

We varied the $T_{\mathrm{eff}}$ and $\log g$ of the synthetic spectra, applying various values for the rotational broadening around the literature value of $v \sin i = 27$\,\kms, and allowing for a radial velocity (RV) offset to fit the ESPaDOnS spectra.  This resulted in a best fit with  $T_{\mathrm{eff}}=15000$\,K,  $\log g=3.8$\,dex, and $v \sin i = 35$\,\kms.  The wings of the Balmer lines are generally well described by the model, while the depths of the \ion{He}{I} lines or several metal lines, such as those of \ion{Mg}{II}, are overestimated.  Peculiar surface abundances connected to the large-scale magnetic field can produce such a discrepancy since the grid relies on a solar composition for the synthetic data.  We show the ESPaDOnS observations and the best synthetic model in Fig.\,\ref{fig:spec_synthetic}, as well as the residuals to the fit.

However, \OmiLup is a known interferometric binary system.  The contrast ratio derived by \citet{2013MNRAS.436.1694R} implies that about 40\,\% of the flux should originate from the secondary component.  Moreover, the angular separation between the two components of \OmiLup is sufficiently small that both fall within the fiber of modern spectrographs.  However, the secondary has never firmly been detected in spectroscopy.  Employing the \textsc{atlas9} and \textsc{cossam\_simple} synthetic spectra, we investigated whether a binary spectrum describes the observations better than a single star.

We varied $T_{\mathrm{eff}}$ for both components from 13000\,K up to 18000\,K and $\log g$ from 3.5\,dex up to 4.5\,dex, allowing various $v \sin i$ values, light fractions from 0\,\% up to 50\,\%, and relative RV offsets up to 50\,\kms.  The best fit occurs for $T_{\mathrm{eff, 1}} = 17000$\,K, $ T_{\mathrm{eff,2 }} = 14000$\,K, $\log g_1 = 3.9$\,dex, $\log g_2 = 3.8$\,dex, $v \sin i_1 = 50$\,\kms, $v \sin i_2=25$\,\kms, $RV_1 =-5$\,\kms, $RV_2 =+5$\,\kms, and a light fraction of 50\,\%.  This model is indicated in Fig.\,\ref{fig:spec_synthetic}. While the description of the helium and metal lines improved, the fit to the Balmer lines did not necessarily improve.  The binarity has a similar effect on the metal lines as a surface under-abundance, which is often observed for magnetic early-type stars.  Moreover, the light fraction of the secondary is higher than determined from interferometry, i.e., the fitting algorithm tries to obtain a better description for the (weaker) metal lines.  For these reasons, we rejected the more complex model, where both stars contributed equally to the spectroscopic observations, and accepted the simpler model where only one star is visible.  Further, we argue that if both components do contribute, we cannot distinguish between them in the spectroscopy due to their similar spectral types, small RV shifts, and expected chemical peculiarities.  This is in agreement with the results of Sect.\,\ref{sec:mag_zeeman}, where we show that the Zeeman signature spans the full width of the average line profiles.  Therefore, we adopt $T_{\mathrm{eff}}=15000$\,K and $\log g=3.8$\,dex as the starting point for the magnetometric analysis.

Incorrect values for $T_{\mathrm{eff}}$ or $\log g$ do impact the analysis of the longitudinal magnetic field as an inappropriate set of lines will be used in the determination of the average line profile in the magnetometric analysis. However, slight discrepancies between the adopted values and the real ones would only have a negligible impact, since the line depth in the line mask will be adjusted to the observations (see also Sect.\,\ref{sec:mag_zeeman}).

\section{Periodic photometric variability}
\label{sec:photometry}
Magnetic early-type stars often show periodic photometric variability due to co-rotating surface abundance inhomogeneities caused by the large-scale magnetic fields.  Moreover, \citet{2011A+A...536L...6A} discussed the possibility of stellar pulsations in \OmiLup, yet attributed the LPVs to surface abundance inhomogeneities.  To determine the cause of the LPVs, we investigated the BRITE photometry for coherent periodic variability.  We employed an iterative prewhitening approach to determine the frequencies of the periodic variability.  Following the spectroscopic results of Sect.\,\ref{sec:st_params}, we attributed all photometric variability in the BRITE data to the primary component.  However, implications and ambiguity caused by the binary system of \OmiLup are further discussed in Sect.\,\ref{sec:discussion}.

\begin{figure*}[t]
		\centering
			\includegraphics[width=\textwidth, height = 0.50\textheight]{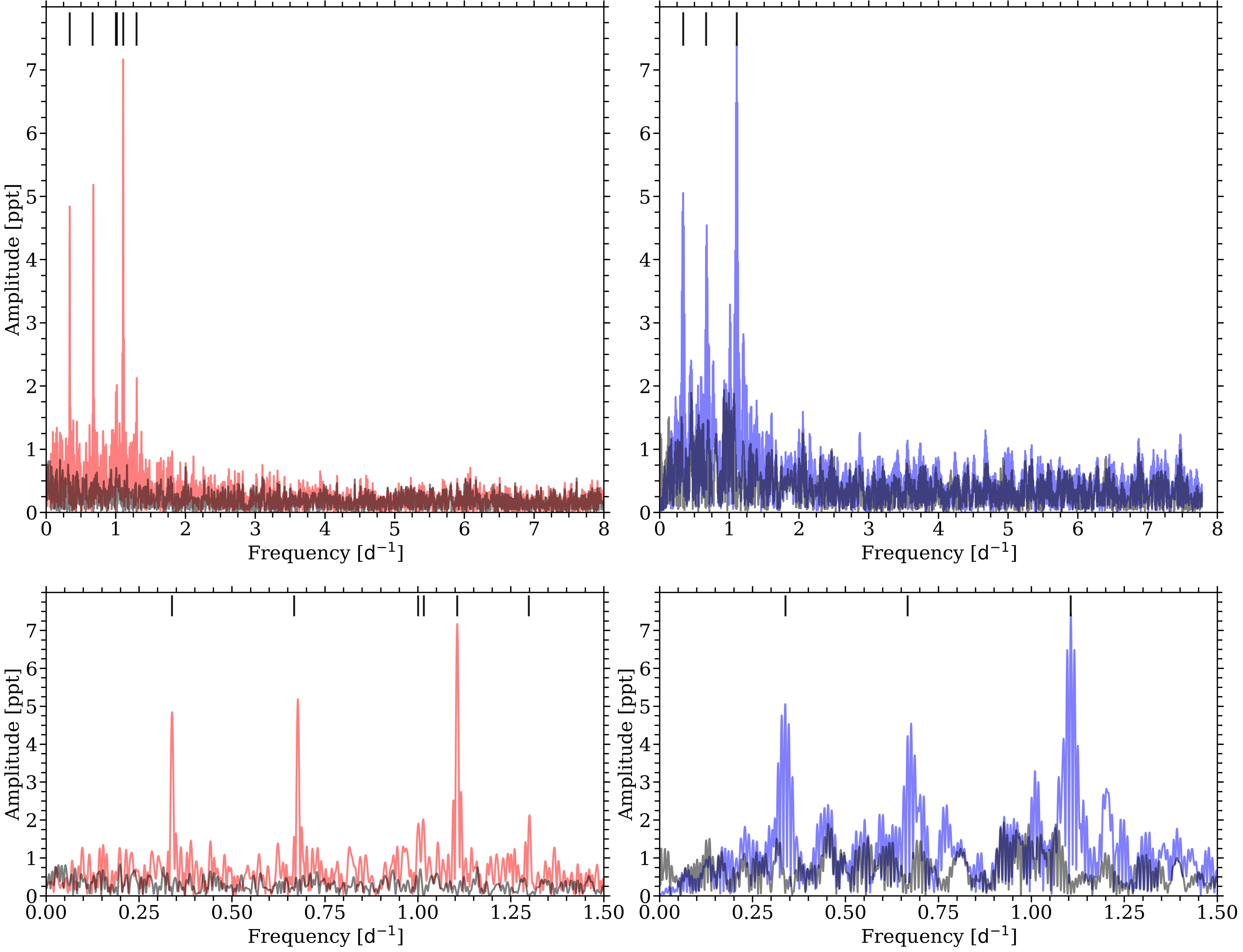}%
			\caption{Lomb-Scargle periodograms showing the periodic variability for the UBr (\textit{left}) and BAb (\textit{right}) light curves.  \textit{Top}: Periodograms covering the full investigated frequency domain.  The periodograms of the light curve are given in red (UBr; \textit{left}) and blue (BAb; \textit{right}), while the variability of the residuals after the iterative prewhitening is given in black.  \textit{Bottom}: Periodograms covering the frequency domain where significant periodic variability is recovered.  The amplitude of the periodic variability is given in parts-per-thousand (ppt). The frequencies of the extracted variability is given by the small black ticks in the top part of each panel.}
			\label{fig:BRITE_periodogram}
\end{figure*}

\begin{table*}[t]
\caption{Significant periodic photometric variability seen in the UBr and BAb BRITE photometry of \OmiLup.}
\centering
\tabcolsep=6pt
\begin{tabular}{ll|cccc|cccc}
\hline
\hline
ID		& Origin						&  \multicolumn{4}{c|}{UBr} 					& \multicolumn{4}{c}{BAb} 					\\
		&							&	$f$ 			& $\delta f$ 		&	$A$ 					& S/N			&	$f$ 		&  $\delta f$ 		&	$A$					& S/N		\\
		&							&	(\d)			& ($10^{-4}$\,\d)	&	($\pm 0.17$\,ppt)	&				& (\d)		&  ($10^{-4}$\,\d)	&	($\pm 0.35$\,ppt)	&			\\
\hline
$f_1$	& $f_{\mathrm{rot}}$			& 0.33869		&	1.5			& 4.63			& 9.6	&	0.33792			&	2.9	& 5.02			& 4.7	\\
$f_2$	& $2f_{\mathrm{rot}}$		& 0.67706		&	1.3			& 5.03			& 8.5	&   0.67687			&	3.2	& 4.62			& 5.0	\\
$f_3$	&	g mode					& 1.10572		&	0.9			& 7.14			& 10.0	&   1.10611			&	1.9	& 7.65			& 5.8	\\
$f_4$	& $f_{\mathrm{inst}}$		& 1.00078		&	3.5			& 1.91			& 5.9	&					&		&				&		\\
$f_5$	& $3f_{\mathrm{rot}}$		& 1.01586		&	5.2			& 1.29			& 4.5	&					&		&				&		\\
$f_6$	&	g mode					& 1.29852		&	4.8			& 1.40			& 4.8	&					&		&				&		\\
\hline
\end{tabular}
\label{tab:BRITE_frequency}
\tablefoot{We indicate the frequency and corresponding amplitude, A, together with their respective uncertainties, as well as the S/N of the detection in the Lomb-Scargle periodogram during the iterative prewhitening procedure.  The frequency and amplitude uncertainties are determined from \citet{1999DSSN...13...28M}, under the assumption of white noise and uncorrelated data.  These conditions are not always fullfilled and result in a typical underestimation of the frequency error by a factor 10 \citep[e.g.,][]{2009A+A...506..471D}. We also indicate the proposed origin of the observed periodic photometric variability and remark that a more precise value for the rotation period was obtained through the magnetometric analysis in Sect.\,\ref{sec:magnetic}.}
\end{table*}

Iterative prewhitening is typically applied to recover and study the stellar pulsation mode frequencies of massive and early-type stars in both ground-based and space-based photometry \citep[e.g.][]{2009A+A...506..111D}.  The method determines the most significant periodic variability, fits a (sinusoidal) model to the data with that frequency, calculates the residuals to the model, and iteratively continues this scheme to the residuals until no significant periodic variability remains.  Adopting this approach, we searched for the significant frequencies in ten times oversampled Lomb-Scargle periodograms \citep{1976Ap+SS..39..447L, 1982ApJ...263..835S} of the BRITE photometry within the 0\,--\,8\,\d frequency range.  No variability was expected at higher frequencies.  The significance of frequency peaks was calculated using the signal-to-noise (S/N) criterion \citep{1993A+A...271..482B} with a frequency window of 1\,\d centered at the frequency of the variability and this after its extraction.  Frequency peaks were considered significant if their S/N reached the threshold value of four.  The periodograms for each BRITE light curve are shown in Fig.\,\ref{fig:BRITE_periodogram}.

This method resulted in six significant frequencies for the UBr photometry and three frequencies for the BAb photometry, in the frequency domain of 0\,--\,1.5\,\d.  We report these in Table\,\ref{tab:BRITE_frequency}, together with their respective uncertainties.  It was expected that the analysis of the BAb would result in less clear periodic variability, represented by the smaller amount of significant frequencies, due to the large time gap, which complicated the analysis.  We note that $f_2$ is the second frequency harmonic of $f_1$ and $f_5$ is the third harmonic of $f_1$, making it very likely that these three frequencies are related to the rotational modulation of the magnetic component.  (We confirm this hypothesis in Sect.\,\ref{sec:magnetic}.)  Moreover, $f_4$ is very close to 1.0\,\d, which is a known instrumental frequency for the BRITE photometry, related to the periodic on-board temperature variability \citep[e.g., Fig.\,A.4 of ][]{2017A+A...602A..91B}.  No significant amplitude changes were retrieved during the length of the BRITE light curves.

Periodic variability with the same frequency in both the UBr and BAb photometry had comparable amplitudes (see Table\,\ref{tab:BRITE_frequency}).  The frequencies $f_3$ and $f_6$ are likely due to a g-mode pulsations.  In this case, the slightly higher amplitude for $f_3$ in the blue filter was expected.  Yet, without additional and simultaneous time resolved photometry employing different bandpass filters, performing mode identification of the stellar pulsations with the amplitude ratio method was impossible \citep[see e.g.,][where this method was successfully applied for a pulsating early-type star with BRITE and ground-based photometry]{2017MNRAS.464.2249H}.  As the retrieved variability had similar amplitudes, we ignored the colour information of the individual BRITE light curves and combined these into one light curve.  This was done once without any weighting methods and once with a simple weighting method (using the $\mathrm{rms_{corr}}$ values) to account for differences in data quality, but overall no simple weighting method is available for BRITE photometry \citep[see also][for more general information on possible weighting methods]{2003BaltA..12..253H}.  Subsequent iterative prewhitening did not result in any new significant periodic variability compared to what was already obtained from the UBr data.

Once all significant periodic photometric variability was subtracted from the UBr photometry, the frequency diagram of the residuals seems to be nearly constant with amplitudes well below 1\,ppt (see Fig.\,\ref{fig:BRITE_periodogram}).  No obvious variability remained in the residual light curves in the time domain.  Moreover, the noise level in the periodogram of the residual BAb photometry was considerably higher than that of the UBr residuals, because it has less data points.

\section{Magnetic measurements}
\label{sec:magnetic}
\subsection{Zeeman signatures}
\label{sec:mag_zeeman}
\begin{figure*}[t]
		\centering
			\includegraphics[width=\textwidth, height = 0.33\textheight]{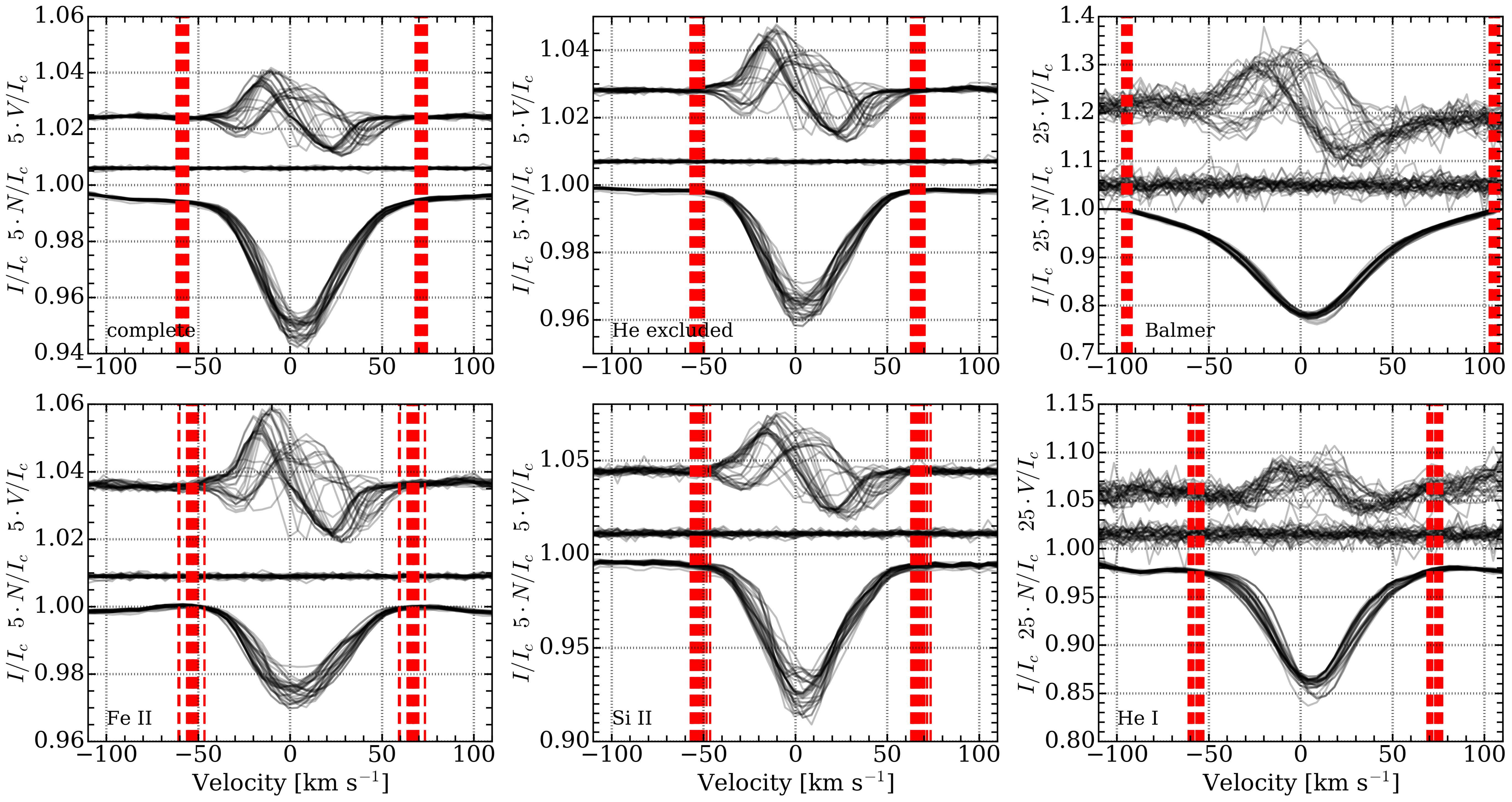}%
			\caption{Overplotted mean line profiles constructed with the LSD method using various line masks.  Each panel shows (from top to bottom) the LSD Stokes\,V profile, the diagnostic null profile, and the LSD Stokes\,I profile for the observations, that are off-set for increased visibility.  Differences in the line depth of the LSD Stokes\,I profile, the strength and shape of the Zeeman signature in the LSD Stokes\,V profile, and the shape of the LPVs in the LSD Stokes\,I profile are clearly visible.  We indicate the integration limits for the computation of the longitudinal magnetic field and the FAP determination by the red dashed lines for each observation.  For the Balmer lines, only the cores of the LSD Stokes\,I profiles were used and shifted upwards to unity, following the technique of \citet[][see text]{2015A+A...580A.120L}.}
			\label{fig:LSD_overlay}
\end{figure*}

To reliably detect the Zeeman signature of a stable large-scale magnetic field at the surface of an early-type star, mean line profiles are constructed from each high-resolution spectropolarimetric observation to boost the S/N of the signature in the Stokes\,V polarization.  We employed the Least-Squares Deconvolution (LSD) technique \citep{1997MNRAS.291..658D} to create these mean line profiles.  We started from a pre-computed \textsc{vald3} line mask \citep{2015PhyS...90e4005R} with $T_{\mathrm{eff}}=15000$\,K and $\log g = 4.0$\,dex, which is the closest match to our results $T_{\mathrm{eff}}=15000$\,K and $\log g = 3.8$\,dex from the simpler model describing the spectroscopic observations (see Sect.\,\ref{sec:st_params} for a discussion between both used models).  We removed from the line mask all hydrogen lines and all helium and metal lines that were blended with hydrogen lines, telluric features, and known diffuse interstellar bands to ensure we only included absorption lines with similar line profiles (the only exception being the \ion{He}{I} lines where pressure broadening through the Stark effect is still significant).  All metal lines with a depth smaller than 0.01 were also discarded.  Lastly, the depths of the lines included in the line mask were adjusted to correspond to the observations \citep[a technique sometimes referred to as "tweaking the line mask", see e.g.,][]{2017MNRAS.465.2432G}.  This resulted in a final line mask with 893 lines included, each with their respective wavelength, line depth, and Land{\'e} factor.  An additional line mask (and corresponding LSD profiles) without (blends with) \ion{He}{I} lines was also constructed and included 821 metal lines.  We show these mean line profiles in Fig.\,\ref{fig:LSD_overlay} for both line masks.  The Zeeman signature in the LSD Stokes\,V profile clearly varies between the various observations.  Moreover, the LSD Stokes\,I profile exhibits clear LPVs, possibly due to the rotational modulation caused by surface abundance inhomogeneities or the stellar pulsation.  Fortunately, the diagnostic null profiles \citep[i.e., deconstructively added polarization signal within the sequence,][]{1997MNRAS.291..658D} suggest that no significant instrumental effects or LPVs have occurred during the spectropolarimetric sequence, as they are flat.

To determine whether an LSD Stokes\,V spectrum contained a Zeeman signature, we determined the False Alarm Probability \citep[FAP;][]{1992A+A...265..669D, 1997MNRAS.291..658D} for each LSD profile.  Definitive detections (DD) show a clear signature and correspond to a FAP\,$ < 10^{-3}\,\%$, while non-detections (ND) have a FAP\,$ > 10^{-1}\,\%$.  Marginal detections (MD) fall in between DDs and NDs, with $10^{-3}\,\% < $\,FAP\,$ < 10^{-1}\,\%$.  Based on these criteria, all LSD profiles indicated a DD, irrespective whether the complete line mask or the He-excluded line mask was used.  Thus, a large-scale magnetic field is clearly detected in the high-resolution spectropolarimetry.

Figure\,\ref{fig:LSD_overlay} further indicates that the Zeeman signature in LSD Stokes\,V spans the full width of the LSD Stokes\,I profile, demonstrating that this full width corresponds to the magnetic component.  No additional spectroscopic component is identified in the LSD Stokes\,I profiles.  Since the interferometric results of \citet{2013MNRAS.436.1694R} indicated that \OmiLup is a binary system with a comparable light fraction, both components must have a similar spectral type and have small RV shifts.  This, together with the presence of LPVs, makes it impossible to discern which component hosts the detected large-scale magnetic field.  These results are in agreement with those from the spectroscopic analysis of Sect.\,\ref{sec:st_params}. As such, we will likely underestimate the true strength of the magnetic field through an overestimated depth of the Stokes\,I profile in Eq.\,(\ref{eq:longfield}).

\begin{table*}[t]
\caption{Parameters related to the study of the longitudinal field measurements from various LSD line masks.}
\centering
\tabcolsep=6pt
\begin{tabular}{llcccccc}
\hline
\hline
\multicolumn{2}{l}{Parameter}& Complete 	&He excluded& Balmer & \ion{Fe}{II}	& \ion{Si}{II} & \ion{He}{I}\\
\hline
$g$				&			&$	1.1906		$&$	1.1877		$&$	1.5000		$&$	1.1939		$&$	1.1446		$&$	1.2474	$\\
$\lambda$		&	[nm]		&$	505.99		$&$	508.79		$&$	486.13		$&$	516.79		$&$	508.46		$&$	478.18	$\\
$\langle \mathrm{EW} \rangle$	&	[\kms]	&$	2.96\pm0.10	$&$	1.83\pm 0.11	$&$				$&$	1.34\pm0.10	$&$	3.41\pm0.30	$&$	8.74\pm0.21	$\\
int. range		&	[\kms]	&${	\pm 65		}$&${	\pm 60		}$&${	\pm 100		}$&${	\pm 60		}$&${	\pm 60		}$&${	\pm 65		}$\\
$B_0$			&	[G]		&${	344 \pm 6	}$&${	821 \pm 11	}$&${	531 \pm 32	}$&${	1679 \pm 31	}$&${	724 \pm 22	}$&${	27 \pm 9		}$\\
$B_1$			&	[G]		&${	568 \pm 9	}$&${	1340  \pm 15	}$&${	913 \pm 45	}$&${	2905 \pm 43	}$&${	1053 \pm 30	}$&${	82 \pm 14	}$\\
$B_2$			&	[G]		&${	61 \pm 8		}$&${	168 \pm 13	}$&${	64 \pm 46	}$&${	383 \pm 37	}$&${	57 \pm 29	}$&${	1 \pm 12		}$\\
$\phi_1$			&	 		&${	0.719 \pm 0.002	}$&${	0.722 \pm 0.002 }$&${	0.741 \pm 0.007 	}$&${	0.722 \pm 0.002 }$&${	0.735 \pm 0.004		}$&${	0.676 \pm 0.022	}$\\
$\phi_2$			&	 		&${	0.46 \pm 0.02	}$&${	0.45 \pm 0.01  	}$&${	0.77 \pm 0.10 	}$&${	0.43 \pm 0.01  	}$&${	0.36 \pm 0.08		}$&${	0.70 \pm 0.14	}$\\
\hline
\end{tabular}
\label{tab:LSD_generalinfo}
\tablefoot{We provide the mean Land{\'e} factor $g$ and the mean wavelength $\lambda$ from the LSD calculation, as well as the resulting mean equivalent width (EW) of the LSD Stokes I profiles.  The integration range around the line centroid for the calculation of the longitudinal magnetic field (see Eq.\,(\ref{eq:longfield})) and determination of the FAP is given, as well as the resulting parameters of the dipole with a quadrupole contribution model (i.e., Eq.\,(\ref{eq:dipolequad_modulation})) fitted to the measured longitudinal field.  The detection status following the computed FAP value for a specific observation and LSD line mask is provided in Table\,\ref{tab:specpol_log}, when not all observations had a definite detection for that specific line mask.}
\end{table*}

\subsection{Longitudinal field measurements}
\label{sec:mag_longmag}
Since the large-scale magnetic fields of early-type stars are expected to be stable over long time scales and inclined with respect to the rotation axis, the measured longitudinal magnetic field should exhibit rotational modulation (depending on the relative orientation to the observer). This rotational modulation can be used to accurately determine the rotation period of the magnetic component of \OmiLup.

The longitudinal magnetic field \citep[in Gauss, see][]{1979A+A....74....1R} is measured as 
\begin{equation}
B_l = -2.14 \cdot 10^{11} \frac{\int v V(v) \mathrm{d}v}{\lambda g c \int[1-I(v)] \mathrm{d}v} \, \mathrm{,}
\label{eq:longfield}
\end{equation}
\noindent where $V(v)$ and $I(v)$ are the LSD Stokes\,V and I profiles for a given velocity $v$.  The parameters $g$, the mean Land{\'e} factor, and $\lambda$, the mean wavelength (in nm) come from the LSD method.  The speed of light is given by $c$ (in \kms).  We provide $g$ and $\lambda$ for the various line masks in Table\,\ref{tab:LSD_generalinfo} and the determined values for $B_l$ in Table\,\ref{tab:appendix_Bl_values}.  The integration limits to determine the $B_l$ should cover the full LSD Stokes\,I profile, and thus, also the full Zeeman signature.  Following the plateau method \citep[e.g., Fig.\,3 of][]{2012A+A...537A.148N}, where we investigated the dependency of $B_l$ and $\sigma(B_l)$ with the integration limit for a near magnetic pole-on observation, we determined an integration range of $\pm65$\,\kms and $\pm60$\,\kms around the line centroid to be the most appropriate for the complete and the He-excluded LSD profiles, respectively.  These values were considerably larger than the literature $v\sin i = 27$\,\kms for \OmiLup, which did not capture the complete width of the absorption profile or the Zeeman signature due to the application of the LSD technique.

When the large-scale magnetic field has a pure dipolar geometry, the rotational modulation of the measured longitudinal magnetic field can be characterized by a sine model:
\begin{equation}
B_l(t) = B_0 + B_1 \sin\left(2\pi \left(f_{\mathrm{rot}} t + \phi_1\right)\right)  \, \mathrm{,}
\label{eq:dipole_modulation}
\end{equation}
\noindent where $B_1$ and $\phi_1$ are the amplitude and phase of the sine, $B_0$ the constant offset, and $f_{\mathrm{rot}}$ the rotation frequency.  However, when the large-scale magnetic field has a dipolar component with a non-negligible quadrupolar contribution, the longitudinal magnetic field modulation is given by a second-order sine model:
\begin{equation}
B_l(t) = B_0 + B_1 \sin\left(2\pi \left(f_{\mathrm{rot}} t + \phi_1\right)\right) + B_2 \sin\left(2\pi \left(2f_{\mathrm{rot}} t + \phi_2\right)\right)  \, \mathrm{.}
\label{eq:dipolequad_modulation}
\end{equation}
\noindent Again, $B_i$ and $\phi_i$ are the amplitude and phase of the individual sine terms.  We fitted both models to the longitudinal magnetic field measurements of the He-excluded LSD profiles using a Bayesian Markov Chain Monte Carlo (MCMC) method \citep[using \textsc{emcee};][]{2013PASP..125..306F} to determine the rotation period.  We adopted the log-likelihood function for a weighted normal distribution:
\begin{multline}
\Lagr(\Theta) = -\frac{1}{2} N \ln \left(2 \pi \right) - N \sum\limits_{i=1}^{N}\ln \left( \sigma \left(B_l(t_i) \right)\right) \\ - \sum\limits_{i=1}^{N}\left( \frac{\left(B_l(t_i) - \Magr(\Theta; t_i)\right)^2}{2 \sigma \left(B_l(t_i) \right)^2}\right) \ \mathrm{,}
\label{eq:loglikelihood}
\end{multline}
\noindent with $\ln$ the natural logarithm, $\Magr(\Theta; t_i)$ the model of Eq.\,(\ref{eq:dipole_modulation}) or Eq.\,(\ref{eq:dipolequad_modulation}) for a given parameter vector $\Theta$ at timestep $t_i$.  The measured longitudinal magnetic field at $t_i$ is given by $B_l(t_i)$, its respective error by $\sigma \left(B_l(t_i) \right)$, and $N$ is the number of observations.  We constructed uniform priors in the appropriate parameter spaces for each free parameter describing the models of Eq.\,(\ref{eq:dipole_modulation}) and Eq.\,(\ref{eq:dipolequad_modulation}).  For the rotation frequency, this was around $f_1$ from the BRITE photometry with a range set by the Rayleigh frequency criterion employing the time length of the spectropolarimetric dataset ($f_{\mathrm{res}} = 0.0006$\,\d).  Calculations were started at random points within the uniform parameter space employing 128 parameter chains and continued until stable frequency solutions were reached.

\begin{figure}[t]
		\centering
			\includegraphics[width=0.45\textwidth, height = 0.4\textheight]{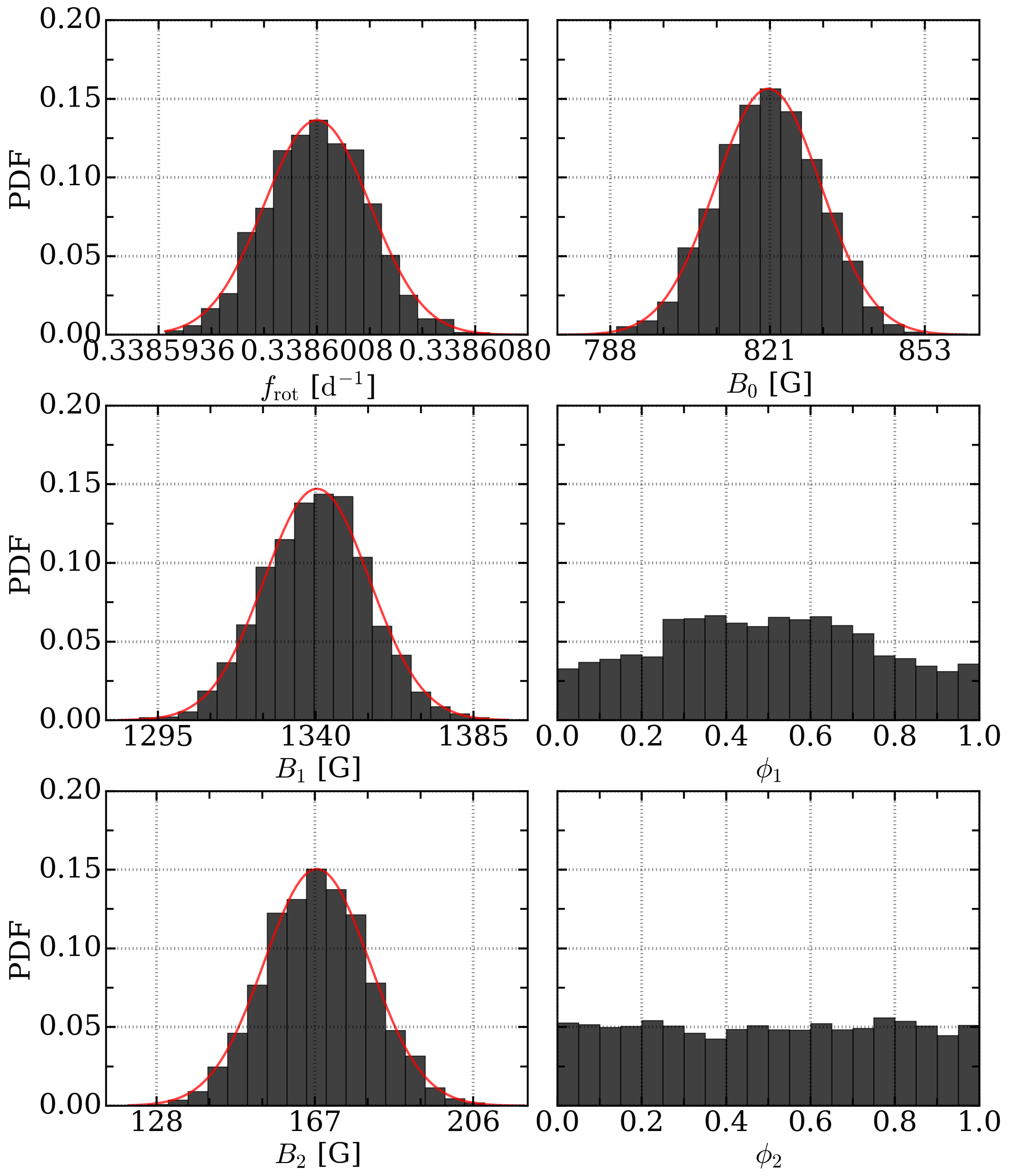}%
			\caption{Results of the MCMC analysis on the measured longitudinal magnetic field of the He-excluded LSD profiles using the model of Eq.\,(\ref{eq:dipolequad_modulation}).  The posterior probability distributions (PDF) are indicated for each fitted parameter.  When these had a normal distribution, we represent it with the Gaussian description marked in red.}
			\label{fig:longmag_noHe_MCMC}
\end{figure}

\begin{figure}[t]
		\centering
			\includegraphics[width=0.45\textwidth, height = 0.66\textheight]{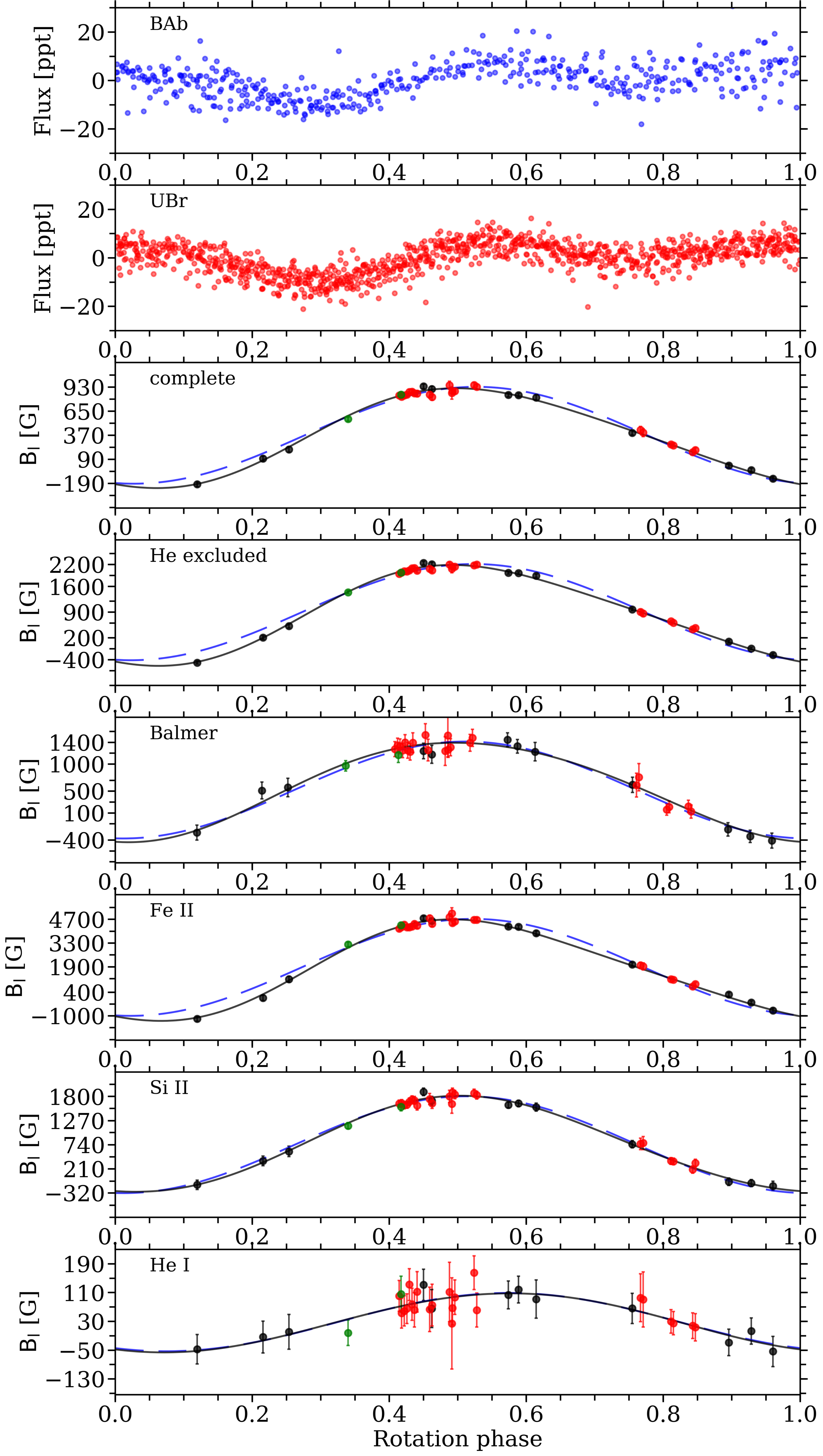}%
			\caption{Phase folded BRITE light curves (after subtracting the non-rotational variability); \textit{upper two panels}) and longitudinal magnetic field measurements from LSD profiles with various line masks, using a $P_{\rm rot} = 2.95333$\,d and $T_0 = \mathrm{HJD}\,2455702.5$, and indicated by the observing campaign (black for MiMeS, red for BRITEpol, and green for ESPaDOnS).  The pure dipole model for the geometry of the large-scale magnetic field of Eq.\,(\ref{eq:dipole_modulation}) is indicated by the dashed blue line and the dipole with a quadrupole contribution of Eq.\,(\ref{eq:dipolequad_modulation}) is indicated by the solid black line.  Note the differences in strength of the measured $B_l$ values for the various LSD line masks.}
			\label{fig:longmag_rotation}
\end{figure}

The posterior probability distribution function (PDF; see Fig.\,\ref{fig:longmag_noHe_MCMC}) of the rotation frequency of Eq.\,(\ref{eq:dipolequad_modulation}) showed a maximum at $f_{\mathrm{rot}} = 0.338601(2)$\,\d, corresponding to a rotation period of $P_{\mathrm{rot}} = 2.95333(2)$\,d.  Also the PDF for the amplitudes $B_1$ and $B_2$, and the offset $B_0$ show a normal distribution centered at a non-zero value.  The distributions for the phases $\phi_1$ and $\phi_2$ are nearly uniform, which is expected as $f_{\mathrm{rot}}$ was a free parameter during this process and each value of $f_{\mathrm{rot}}$ from the PDF creates a normal distribution for the phase $\phi_i$.  The superposition of these normal distributions for $\phi_i$ creates the obtained nearly uniform distribution.  We indicate the recovered PDFs for the fitting parameters in Fig.\,\ref{fig:longmag_noHe_MCMC} and list the deduced values for $B_0$, $B_1$, and $B_2$, with their respective uncertainties, in Table\,\ref{tab:LSD_generalinfo}.  The fit to the $B_l$ values computed from the He excluded LSD profile resulted in a non-zero value for $B_2$, indicating that the second-order term of the model is needed, hence the quadrupolar component of the large-scale magnetic field is significant.  This was further supported by the information criteria during the fitting process and model selection.  Once a value for each fitted parameter was obtained, we defined an initial epoch $T_0$ around the first observations (i.e., H01), to place the maximum of the (sine-model) at a rotation phase of 0.5.

Assuming that both components of \OmiLup contribute to the spectropolarimetric data and that their rotation periods are significantly shorter than the length of the spectropolarimetric time series, we assert that only one component hosts a strong large-scale magnetic field.  Indeed, there is a lack of variability in the measured longitudinal magnetic field other than with a period $P_{\mathrm{rot}}$.  A second magnetic component would severely distort the rotational modulation indicated in Fig.\,\ref{fig:longmag_rotation}.  Only a binary system with synchronized rotation periods could reproduce this variability, which is highly unlikely for the \OmiLup binary system given its $P_{\mathrm{orb}} > 20$\,years.

We included the BRITE light curves, phase folded with $P_{\mathrm{rot}}$ and the periodic variability not caused by the rotation removed, in Fig.\,\ref{fig:longmag_rotation} to compare the photometric rotational modulation with that of the longitudinal magnetic field.  The peak brightness in the folded BRITE light curves occurs close to the phase where we observed the magnetic poles (at rotation phases 0.0 and 0.5), suggesting that the brighter surface abundance inhomogeneities are located close to the magnetic poles.  The remaining photometric variability depends on the inclination angle $i$, the obliquity angle $\beta$, and the relative positions and sizes of the abundance inhomogeneities at the stellar surface, which requires tomographic imaging and/or spot modelling and is beyond reach of the current data.

\subsection{Single-element longitudinal field measurements}
\label{sec:mag_singlelongmag}

While adjusting the line depths of the metal lines used in the LSD line masks, we noted that the Zeeman signature shows different strengths for different chemical species.  Therefore, we constructed three different line masks containing only 19 \ion{He}{I}, 264 \ion{Fe}{II}, or 61 \ion{Si}{II} metal lines.  Special care was taken to exclude any line blends with metal lines that had a stronger or similar line depth than the considered chemical element.  The corresponding mean Land{\'e} factors and mean wavelengths are given in Table\,\ref{tab:LSD_generalinfo}, and the LSD profiles themselves are given in Fig.\,\ref{fig:LSD_overlay}.  Similar to the LSD profiles constructed with the complete line mask, we obtained DDs for all LSD profiles constructed with either \ion{Si}{II} or \ion{Fe}{II} lines.  For the \ion{He}{I} LSD profiles, however, we obtained ten NDs, and 26 DDs, most likely caused by the lower S/N, as fewer lines were included in constructing the mean line profile.  Indeed, the S/N in the LSD Stokes\,V and Stokes\,I profiles of the \ion{He}{I} line mask was typically three times lower than that of the \ion{Fe}{II} LSD profiles (see Table\,\ref{tab:specpol_log}).  Again, we noted strong LPVs for the LSD Stokes\,I profiles of all single-element LSD line masks, while the diagnostic null profiles indicated that no LPVs or instrumental effects have occurred during the spectropolarimetric sequence.  We study these LPVs in more detail in Sect.\,\ref{sec:LPV}.

We used the single-element LSD line masks to measure the longitudinal magnetic field associated with the large-scale magnetic field.  Again, we employed the plateau method to determine the integration range for the computations.  This resulted in an integration limit of $65$\,\kms, $60$\,\kms, and $60$\,\kms around the line centroid for the \ion{He}{I}, \ion{Fe}{II}, and \ion{Si}{II} line masks, respectively.  Keeping the rotation frequency fixed to the previously derived value, we performed a Bayesian MCMC fit to model the rotational modulation of the measured longitudinal magnetic field.  These models and the measured values are shown in Fig.\,\ref{fig:longmag_rotation}.

The rotational modulation of the measured $B_l$ values from the LSD profiles constructed with only \ion{Fe}{II} lines favored the model of a dipolar magnetic field with a quadrupolar contribution.  This is in agreement with the results for the complete LSD line masks derived in Sect.\,\ref{sec:mag_longmag}.  For the analysis of the measured longitudinal magnetic field of \OmiLup with only \ion{Si}{II} lines, we obtained a less clear distinction between both models for the rotational modulation of the measured $B_l$ values.  Both models agree with the measurements, within the derived uncertainties on the $B_l$ values.  Yet, the fit with the model of dipolar magnetic field and a quadrupolar component was preferred from the information criteria.  Lastly, the $B_l$ values measured from the \ion{He}{I} LSD line mask were much smaller.  This caused the rotational modulation of the measured $B_l$ values to be minimal, leading to a comparable description by both models.  Since the deduced errorbars on the measured $B_l$ values for the \ion{He}{I} remained comparable to those derived from the other LSD line masks, it seemed likely that the noted differences are (astro)physical.  Surface abundance inhomogeneities structured to the geometry of the large-scale magnetic field at the stellar surface could be an explanation.  Elements that are concentrated at the magnetic poles will lead to larger $B_l$ values, while elements located close to the magnetic equator will result in smaller $B_l$ values.  Such features should be noted during tomographic analyses (i.e., Zeeman Doppler Imaging; ZDI), but require a spectropolarimetric dataset which is more evenly sampled over the rotation period than the current observations.

Finally, we tried to model the Zeeman signature of the large-scale magnetic field, seen in the LSD Stokes\,V profile, using a grid-based approach \citep[see e.g.,][for further details]{2008MNRAS.385..391A}.  However, we were not able to accurately model the changing Zeeman profile with varying rotation phase.  This was likely caused by the insufficient sampling of the rotation phase at key phases.

\subsection{Balmer lines longitudinal field measurements}
\label{sec:mag_balmer}
The magnetometric analysis of single-element LSD profiles exhibited a strong scatter in the strength of the measured $B_l$ values, suggesting surface abundance inhomogeneities for certain chemical elements (e.g., He, Si, and Fe).  To measure the rotational modulation of the longitudinal magnetic field for an element that should be homogeneously distributed over the stellar surface we analyzed hydrogen lines.  The wavelength regions around the Balmer lines in the spectropolarimetric observations were normalized with additional care, employing only linear polynomials, so as not to influence the depth of the line core or the broad wings.

We constructed a mean line profile for the Balmer lines, including H$\alpha$, H$\beta$, and H$\gamma$ in the LSD line mask.  Three of these profiles indicated a ND and one a MD of a Zeeman signature in the observations, most likely due to the lower S/N in the Stokes\,V profiles for these observations.  We then followed the method of \citet{2015A+A...580A.120L} to measure the $B_l$ values.  This method uses only the core of the line and ignores the broad wings.  Moreover, to scale the measurements more in line with those from the metal lines, the (LSD) Stokes\,I profile was not integrated from unity, but instead from the intensity level, $I_c$, where the core transitions into the wings.  As the Zeeman signature in the LSD Stokes\,V profile is slightly wider than the core of the Stokes\,I profile, we employed this width to set the integration range to 100\,\kms around the Stokes\,I line centroid.   We present these LSD profiles in Fig.\,\ref{fig:LSD_overlay}, where the indicated Stokes\,I profile is shifted upwards to place $I_c$ at unity.   While fixing the rotation frequency, we performed a Bayesian MCMC fit to determine the fitting parameters for the description of the rotational modulation of the measured $B_l$ values.  The fit of both models to the measured $B_l$ values is given in Fig.\,\ref{fig:longmag_rotation}, with the parameters in Table\,\ref{tab:LSD_generalinfo}.
 
The rotational modulation of the measured longitudinal magnetic field from the Balmer lines was more accurately represented by the model for a dipolar magnetic field and a quadrupolar component.  This result agreed with those of the other LSD profiles.  Yet, the discrepancies between this model and that of a dipolar magnetic field were small at most of the rotation phases, due to large uncertainties in $B_l$, caused by the low S/N in the LSD Stokes\,V profiles.

\section{Line profile analysis}
\label{sec:LPV}

\begin{figure*}[t]
		\centering
			\includegraphics[width=\textwidth, height = 0.50\textheight]{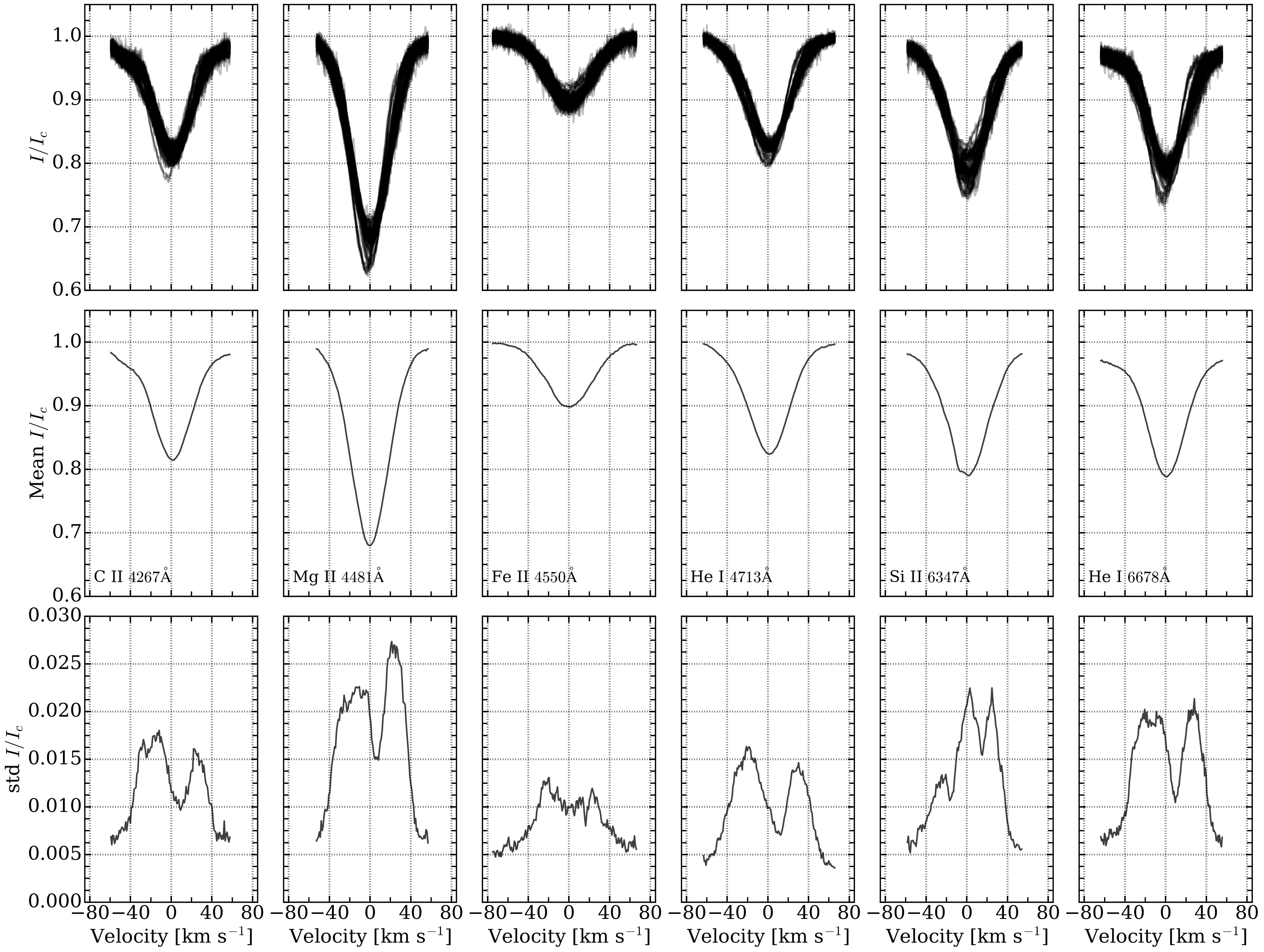}%
			\caption{LPVs for various selected absorption lines.  \textit{Top}: all observations of the given line overlayed to illustrate the LPVs.  As an additional diagnostic, we indicate the mean profile (\textit{middle}) and standard deviation (std; \textit{bottom}) of all observations.  The standard deviation, particularly, shows strong differences when comparing the selected absorption lines.}
			\label{fig:LPV_singlelines}
\end{figure*}

Pulsating early-type stars and magnetic early-type type stars are both known to exhibit LPVs.  \citet{2011A+A...536L...6A} already noted such behaviour for \OmiLup.  Therefore, we investigated the zeroth and first moment of selected absorption lines for periodic variability employing the software package \textsc{famias} \citep{2008CoAst.155...17Z}.  The analysis of six absorption lines, using the sub-exposures of the spectropolarimetric sequences, is presented in Sect.\,\ref{sec:LPV_individual}.  We also investigated the H$\alpha$ line for variability in Sect.\,\ref{sec:LPV_balmer}, to try to diagnose the presence of a magnetosphere.  Lastly, we examined the possibility of co-rotating surface abundance inhomogeneities by analyzing the zeroth moment of six absorption lines in Sect.\,\ref{sec:LPV_spots}.

\subsection{Individual lines}
\label{sec:LPV_individual}
To analyze the LPVs of stellar pulsations in absorption lines, it is preferred to work with deep and unblended absorption lines.  This remains valid when investigating the signatures caused by the rotational modulation of surface abundance inhomogeneities.  During the analysis of the Zeeman signature in the LSD Stokes\,V profile, we noted differences between different chemical species.  Therefore, we selected absorption lines from various elements.  For He, we selected and analyzed the \ion{He}{I}\,4713.2\,$\AA$ and the \ion{He}{I}\,6678.2\,$\AA$ lines, which are close multiplets of several He lines.  In addition, the \ion{Mg}{II}\,4481.1\,$\AA$ fulfilled the set criteria.  At an effective temperature of 15000\,K, there are not many strong and suitable Si lines.  Therefore, we opted for the \ion{Si}{II}\,6347.1\,$\AA$, although it is blended with the weaker \ion{Mg}{II}\,6347.0\,$\AA$ line.  Similarly, we selected the \ion{Fe}{II}\,4549.5\,$\AA$ line, which blends with a weak \ion{Ti}{II} line.  Lastly, we chose the \ion{C}{II}\,4267.3\,$\AA$ line, which is actually a multiplet.  The spectroscopy employed during the LPV analysis was the individual sub-exposures of the spectropolarimetric sequences.  We removed the last two sub-exposures of the observation H25 (see Table\,\ref{tab:specpol_log}), since they did not contain any flux due to bad weather.  This resulted in a total of 142 spectroscopic observations taken over ${\sim}1780$\,days, with large time gaps in between each observing campaign.  We show the selected spectroscopic lines in Fig.\,\ref{fig:LPV_singlelines}, together with their mean line profile and the standard deviation of all observations of a given absorption line.  The latter indicated a different shape for the LPVs for the \ion{Fe}{II} and \ion{Si}{II} lines compared to the other selected absorption lines, such as the \ion{He}{I} or \ion{C}{II} lines.  This might indicate a different dominant variability, and hence a different origin or cause for the LPVs.  The standard deviation of the selected \ion{Mg}{II} line also looked slightly different, yet it seemed to be an intermediate profile between the two studied \ion{He}{I} lines.

\begin{figure*}[t]
		\centering
			\includegraphics[width=\textwidth, height = 0.66\textheight]{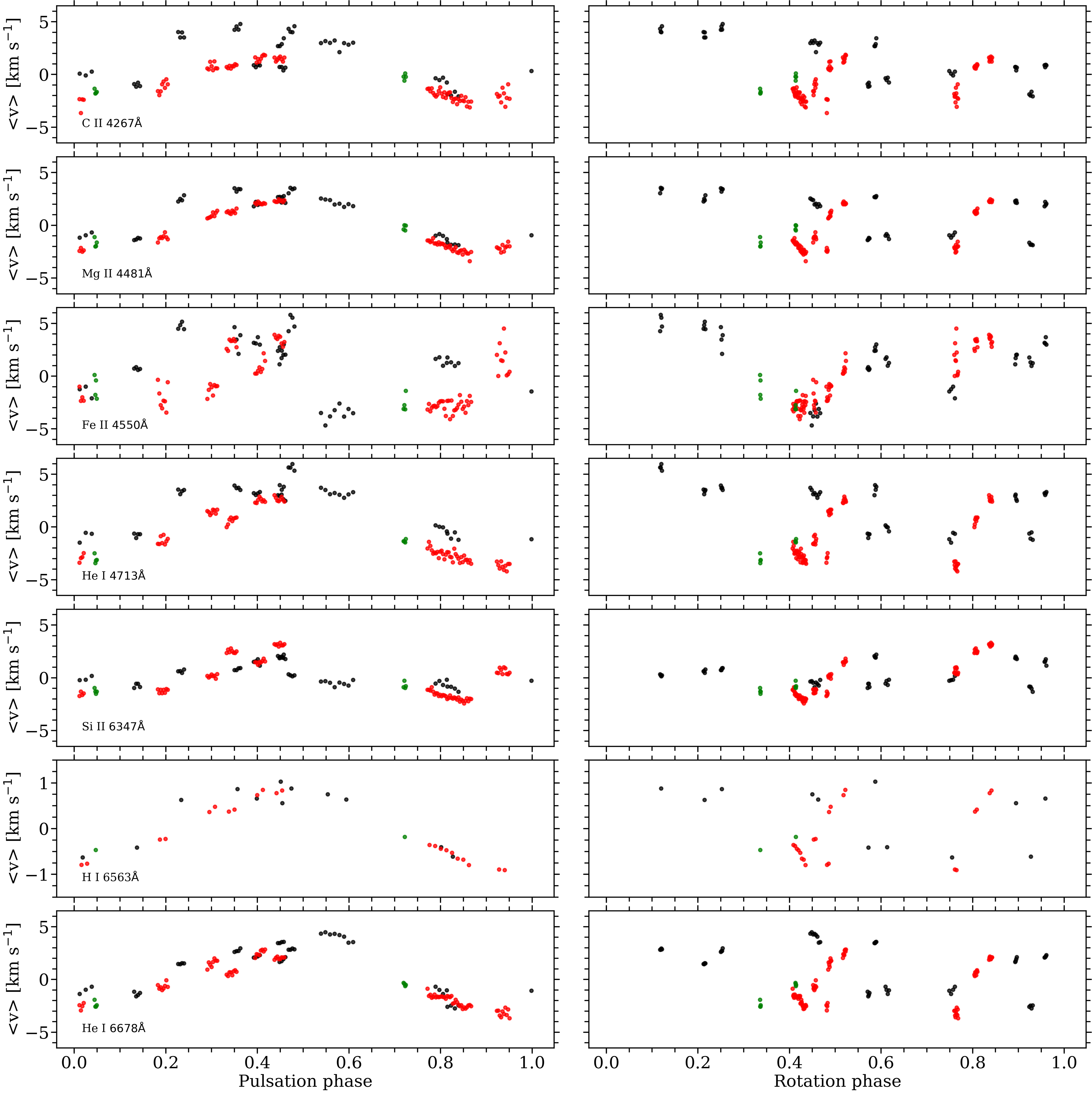}%
			\caption{Phase folded $\langle v \rangle$ for selected absorption lines with the average derived g-mode pulsation frequency ($f_{\rm puls} = 1.10573$\,\d and $T_0 = \mathrm{HJD}\,2455702.0$; see Table\,\ref{tab:LPV_results}); \textit{left} and the rotation period from the magnetometric analysis ($P_{\rm rot} = 2.95333$\,d and $T_0 = \mathrm{HJD}\,2455702.5$; \textit{right}).The same colours were used to indicate the different observations as in Fig.\,\ref{fig:longmag_rotation}.}
			\label{fig:LPV_firstmoment}
\end{figure*}

For each line selected, we set appropriate limits for the calculation of their moments.  These limits were set at similar flux levels for a given spectral line, close to the continuum level, unless strong pressure broadening (e.g.\,\ion{He}{I}\,6678.2\,$\AA$) or asymmetries in the line wings (e.g., \ion{C}{II}\,4267.3\,$\AA$) were noted, resulting in a more narrow range.  We determined the zeroth, $\langle v^0 \rangle$, first, $\langle v \rangle$, and second moment, $\langle v^2 \rangle$, (representative of the equivalent width, radial velocity, and skewness, respectively) for each selected line with the software package \textsc{famias} \citep{2008CoAst.155...17Z}.  Here we continue the discussion of the coherent periodic variability in the first moment.  The zeroth moment is analyzed in Sect.\,\ref{sec:LPV_spots}.  From the BRITE photometry and the rotational modulation of the longitudinal magnetic field, two phenomena were already known to cause periodic variability, each with a distinct period.  These phenomena are rotation modulation (with $P_{\mathrm{rot}} = 2.95333$\,d) and the dominant g-mode pulsation (with $f_3 = 1.1057$\,\d).  We constructed a model that included the (potential) periodic variability in the measured line moments:
\begin{multline}
\langle v \rangle(t) = C + \sum^{2}_{i=1} A_{\mathrm{rot},i} \sin \left(2 \pi \left(i f_{\mathrm{rot}}t + \phi_{\mathrm{rot},i}\right)\right) + \\
 A_{\mathrm{puls}} \sin \left(2 \pi \left(f_{\mathrm{puls}}t + \phi_{\mathrm{puls}}\right)\right) \ \mathrm{,}
\label{eq:multimodel}
\end{multline}
\noindent where $A_{\mathrm{rot},i}$ and $ A_{\mathrm{puls}}$ are the amplitudes of the variability, $\phi_{\mathrm{rot},i}$ and $\phi_{\mathrm{puls}}$ their respective phases, and $C$ a constant off-set.  We deduced each free parameter with a Bayesian MCMC method.  Uniform priors were assumed for all parameters in their appropriate parameter spaces, in particular $f_{\mathrm{puls}}$ had to agree with the conservative result of the BRITE photometry (i.e., $f_3 = 1.1057 \pm 0.0070$\,\d, where the Rayleigh frequency limit was assumed), and $f_{\mathrm{rot}}$ was kept fixed to the value from the magnetometric analysis.  The quality of the fit was determined by the loglikelihood function for a (non-weighted) normal distribution:
\begin{multline}
\Lagr(\Theta) = -\frac{1}{2} N \ln \left(2 \pi \right) - N \ln \left( \sigma \left(\langle v \rangle \right)\right) \\ -  \frac{\sum\limits_{i=1}^{N}\left(\langle v \rangle(t_i) - \Magr(\Theta; t_i)\right)^2}{2 \sigma \left(\langle v \rangle \right)^2} \ \mathrm{,}
\label{eq:loglikelihood_moment}
\end{multline}
\noindent where $\sigma \left(\langle v \rangle \right)$ is the error on all $\langle v \rangle$, of the order of 1\,\kms. Again, 128 parameter chains were used during the MCMC fitting, starting from random positions within the uniform priors, and computations continued until stable solutions were reached.  The computed values for the parameters in Eq.\,(\ref{eq:multimodel}) and $\Lagr(\Theta)$ are given in Table\,\ref{tab:LPV_results}.  Furthermore, we phase-folded $\langle v \rangle$ with the determined $f_{\mathrm{puls}}$ and with $f_{\mathrm{rot}}$ and show these in Fig.\,\ref{fig:LPV_firstmoment}.

The derived amplitudes indicated that the rotational modulation and the dominant g-mode frequency cause the LPVs noted for the absorption lines.  However, the contribution (i.e., the amplitude) of each variability term in $\langle v \rangle$ differed greatly when comparing the different absorption lines.  Variability due to the stellar pulsation was dominant for the investigated \ion{C}{II}, \ion{Mg}{II}, and \ion{He}{I} lines, while the LPVs in the \ion{Fe}{II} line were due to the rotational modulation.  Contributions to the periodic variability of $\langle v \rangle$ for the studied \ion{Si}{II} line were almost equal.  As such, the occurrence of surface abundance inhomogeneities only occurred for that particular chemical element, while LPVs due to the g-mode were always present.  These observed features can also be seen in Fig.\,\ref{fig:LPV_firstmoment}.  Future tomographic analyses, such as ZDI, should indicate the distribution of the surface abundance inhomogeneities more clearly.

We also obtained a value for the pulsation mode frequency $f_{\mathrm{puls}}$ from the Bayesian MCMC fit to each first moment of each absorption line.  Yet, not all obtained values for the g-mode frequency agreed within the same confidence interval, pointing to heteroscedasticity of the first moment measurements.

\begin{table*}[t]
\caption{Values for some of the derived parameters and the loglikelihood for the Bayesian MCMC fits with Eq.\,(\ref{eq:multimodel}) to the moments of the studied absorption lines.}
\centering
\tabcolsep=6pt
\begin{tabular}{llrrrrrrr}
\hline
\hline
\multicolumn{2}{l}{line}				& moment & $C$ & $A_{\mathrm{rot},1}$ & $A_{\mathrm{rot},2}$ & $A_{\mathrm{puls}}$ & $f_{\mathrm{puls}}$	&	$\Lagr(\Theta) $\\
									&&		& [km\,s$^{-1}$]	 & [km\,s$^{-1}$] & [km\,s$^{-1}$]& [km\,s$^{-1}$] & [\d] \\
\hline
\ion{C}{II} 		& 4267.3\,$\AA$	&	$\langle v \rangle$		&$	0.7	\pm	0.2	$&$	1.2	\pm	0.2	$&$	1.7	\pm	0.2	$&$	2.3	\pm	0.2	$&$	1.10577	\pm	0.00002	$&	-159.6	\\
				&				&	$\langle v ^0 \rangle$	&$	10.2	\pm	0.1	$&$	0.7	\pm	0.2	$&$	0.4	\pm	0.2	$&$	0.0	\pm	0.2	$&$		-				$&	-133.1	\\
\ion{Mg}{II} 	& 4481.1\,$\AA$	&	$\langle v \rangle$		&$	0.4	\pm	0.1	$&$	0.6	\pm	0.2	$&$	0.8	\pm	0.2	$&$	2.4	\pm	0.2	$&$	1.10574	\pm	0.00002	$&	-141.4	\\
				&				&	$\langle v ^0 \rangle$	&$	14.9	\pm	0.1	$&$	0.4	\pm	0.2	$&$	0.1	\pm	0.2	$&$	0.0	\pm	0.2	$&$		-				$&	-133.1	\\
\ion{Fe}{II} 	& 4549.5\,$\AA$	&	$\langle v \rangle$		&$	1.7	\pm	0.1	$&$	2.3	\pm	0.2	$&$	2.0	\pm	0.2	$&$	1.0	\pm	0.1	$&$	1.10567	\pm	0.00003	$&	-207.0	\\
				&				&	$\langle v ^0 \rangle$	&$	5.8	\pm	0.1	$&$	0.5	\pm	0.2	$&$	0.3	\pm	0.2	$&$	0.0	\pm	0.2	$&$		-				$&	-134.6	\\
\ion{He}{I} 		& 4713.2\,$\AA$	&	$\langle v \rangle$		&$	0.9	\pm	0.1	$&$	1.4	\pm	0.2	$&$	1.5	\pm	0.2	$&$	2.9	\pm	0.2	$&$	1.10576	\pm	0.00001	$&	-162.3	\\
				&				&	$\langle v ^0 \rangle$	&$	9.2	\pm	0.1	$&$	0.1	\pm	0.2	$&$	0.1	\pm	0.2	$&$	0.0	\pm	0.2	$&$		-				$&	-131.8	\\
\ion{Si}{II} 	& 6347.1\,$\AA$	&	$\langle v \rangle$		&$	0.1	\pm	0.1	$&$	1.1	\pm	0.2	$&$	0.6	\pm	0.2	$&$	1.5	\pm	0.2	$&$	1.10570	\pm	0.00002	$&	-140.6	\\
				&				&	$\langle v ^0 \rangle$	&$	10.8	\pm	0.1	$&$	1.3	\pm	0.2	$&$	0.0	\pm	0.2	$&$	0.0	\pm	0.2	$&$		-				$&	-133.2	\\
\ion{H}{I} 		& 6562.2\,$\AA$	&	$\langle v \rangle$		&$	0.1	\pm	0.3	$&$	0.0	\pm	0.4	$&$	0.0	\pm	0.3	$&$	0.8	\pm	0.3	$&$	1.10575	\pm	0.00009	$&	-33.4	\\
				&				&	$\langle v ^0 \rangle$	&$	68.6	\pm	0.2	$&$	0.1	\pm	0.4	$&$	0.0	\pm	0.3	$&$	0.0	\pm	0.4	$&$		-				$&	-34.8	\\
\ion{He}{I} 		& 6678.2\,$\AA$	&	$\langle v \rangle$		&$	0.1	\pm	0.2	$&$	0.6	\pm	0.2	$&$	0.7	\pm	0.2	$&$	2.9	\pm	0.2	$&$	1.10575	\pm	0.00003	$&	-159.7	\\
				&				&	$\langle v ^0 \rangle$	&$	11.8	\pm	0.1	$&$	0.1	\pm	0.3	$&$	0.4	\pm	0.2	$&$	0.1	\pm	0.3	$&$		-				$&	-133.4	\\
\hline
\end{tabular}
\label{tab:LPV_results}
\tablefoot{Results for the fits to either $\langle v \rangle$ or $\langle v^0\rangle$ of a given absorption line to determine the amplitude of the periodic variability with $f_{\rm rot}$ and with $f_{\rm puls}$.  The former was kept fixed during the analysis.  The amplitudes $A_{\mathrm{rot},i}$ correspond to variability with $i \times f_{\rm rot}$, the amplitude $A_{\rm puls}$ with the g-mode frequency $f_{\rm puls}$ and $C$ is a constant off-set.  No accurate values for $f_{\rm puls}$ were recovered during the fit to $\langle v^0 \rangle$ due to zero $A_{\rm puls}$ and flat PDFs for $f_{\rm puls}$, demonstrating that the model of Eq.\,(\ref{eq:multimodel}) is overfitting these data.}
\end{table*}

\subsection{Balmer lines}
\label{sec:LPV_balmer}
Magnetic early-type stars can host a magnetosphere in their nearby circumstellar environment.  The interactions of wind material with this magnetosphere could, cause rotationally modulated variability in certain spectroscopic lines, with emission features in H$\alpha$ the easiest to identify.  For \OmiLup, we did not observe such emission profiles, however, we did note variability in the core of the H$\alpha$ line.  We thus repeated the analysis of the line moments, where we restricted their computation to the core of the line, fixing the integration limits where the broad line wing starts.

We performed the same analysis as for the other absorption lines for the cores of the H$\alpha$ line of each complete spectropolarimetric sequence. The Bayesian MCMC fit indicated that we did not detect any rotational modulation in the $\langle v \rangle$ of H$\alpha$, because the $A_{\mathrm{rot},i}$ all agreed with zero.  Therefore, we did not identify variability coming from the (potential) magnetosphere.  Similar to the majority of the lines of the previous section, the MCMC fits indicated that the g-mode pulsation frequency is the dominant source for the LPVs.  The significantly lower $\Lagr(\Theta)$ for the fit to the core of H$\alpha$ is the result of using only 36 spectropolarimetric sequences compared to the 142 spectroscopic observations.

\subsection{Surface abundance inhomogeneities}
\label{sec:LPV_spots}

The equivalent width or zeroth moment can be used as a first approximation to follow the change in the surface abundance of chemical species.  This has been employed before to confirm the presence of co-rotating surface abundance inhomogeneities for magnetic stars \citep[e.g.,][]{1991A+AS...89..121M}.  We repeated the analysis of Sect.\,\ref{sec:LPV_individual} by replacing $\langle v \rangle$ by $\langle v^0 \rangle$ in Eq.\,(\ref{eq:multimodel}) and performing the Bayesian MCMC fit to the $\langle v^0 \rangle$ measurements.  The determined amplitudes and 
$\Lagr(\Theta)$ are provided in Table\,\ref{tab:LPV_results}.

\begin{figure}[t]
		\centering
			\includegraphics[width=0.5\textwidth, height = 0.33\textheight]{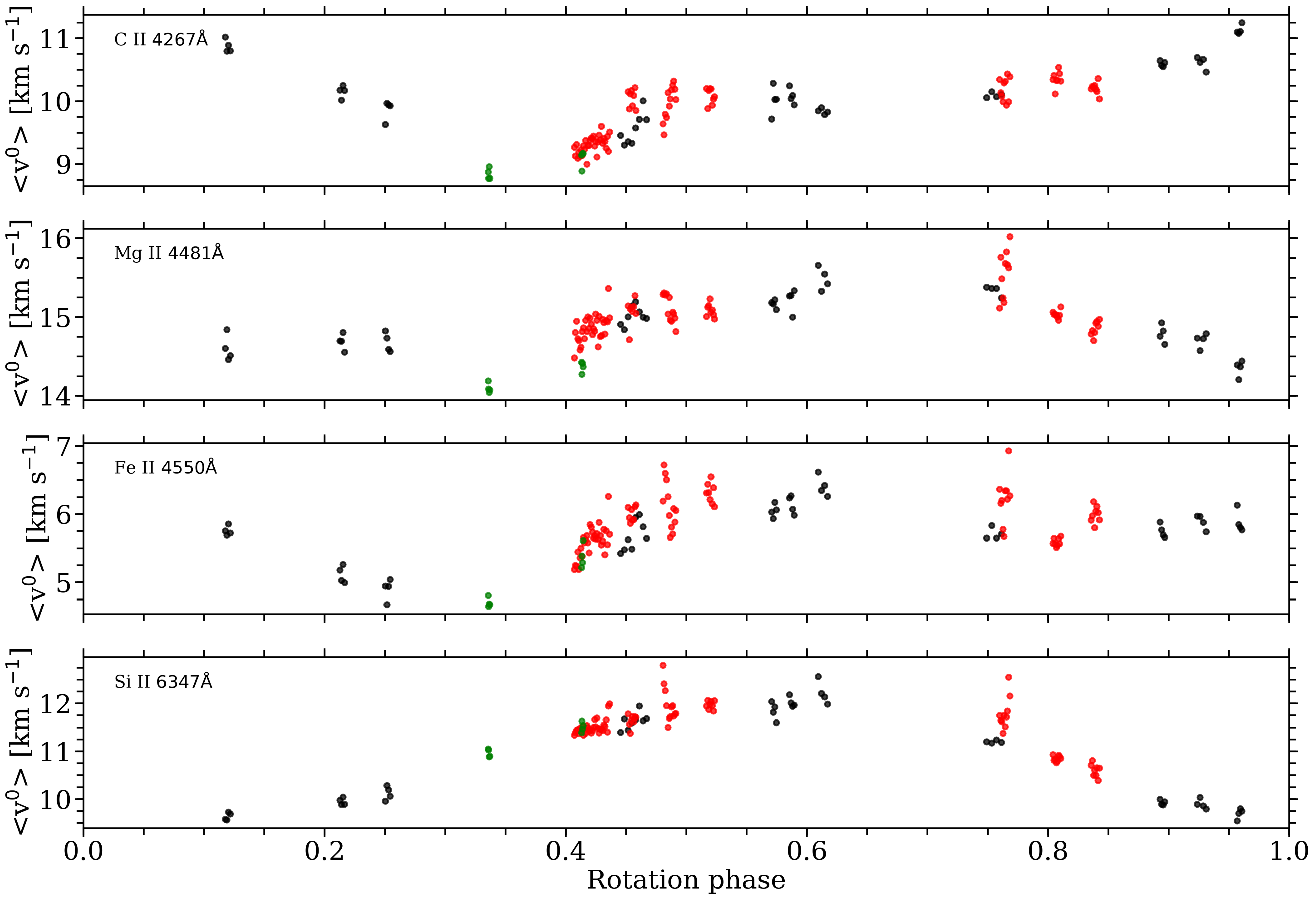}%
			\caption{Phase folded $\langle v^0 \rangle$ for absorption lines, for which non-zero values for $A_{\mathrm{rot},i}$ were derived, with the rotation period ($P_{\rm rot} = 2.95333$\,d and $T_0 = \mathrm{HJD}\,2455702.5$).  The same colours were used to indicate the different observations as in Fig.\,\ref{fig:longmag_rotation}.}
			\label{fig:LPV_zeromoment}
\end{figure}

For each fit to the $\langle v^0 \rangle$ measurements of the different absorption lines, we obtained a value for $A_{\rm puls}$ compatible with zero.  Moreover, the resulting PDFs for $f_{\rm puls}$ were flat.  These results indicated that the g-mode does not cause any significant periodic variability in the $\langle v^0 \rangle$ measurements.  This was expected, as most (non-radial) pulsation modes distort the shape of the line instead of altering the equivalent width.  Furthermore, the fit to the measurements from the \ion{He}{I} lines and H$\alpha$ suggested that their abundances did not vary with the rotation period.  For the remaining four studied lines, some degree of periodic variability with $f_{\rm rot}$ was deduced.  We phase fold the  $\langle v^0 \rangle$ measurements with $f_{\rm rot}$ and show these in Fig.\,\ref{fig:LPV_zeromoment}.

The variability of $\langle v^0 \rangle$ of the \ion{Fe}{II} line can be described by a second order sine function (see also Table\,\ref{tab:LPV_results}), but heavily relies on the scarce measurements between rotation phase 0.20 and 0.35.  Moreover, the phase folded $\langle v^0 \rangle$ of the \ion{Fe}{II} line seems to be coherent with the phase folded BRITE photometry (see top panels of Fig.\,\ref{fig:longmag_rotation}).  The simplest explanation for the LPVs in this Fe line, thus, is the presence of surface abundance inhomogeneities that are located close to the magnetic poles.  Such a geometrical configuration agrees with the stronger measured longitudinal magnetic field from LSD profiles with only Fe lines (Fig.\,\ref{fig:longmag_rotation}) and is often encountered for magnetic Ap/Bp stars \citep[an example for the alignment between the large-scale magnetic field and (He) surface abundance inhomogeneities is presented in][]{2018MNRAS.473.3367O}.

A different profile was obtained for the rotational modulation of the $\langle v^0 \rangle$ of the \ion{Si}{II} line for which only a sinusoid was needed to capture the periodic variability (see Table\,\ref{tab:LPV_results}).  We recall that the amplitudes of the variability caused by the g-mode frequency and the rotational modulation of $\langle v \rangle$ measurements for this \ion{Si}{II} line were comparable.  These results indicate that the Si surface abundance inhomogeneities have a different location on the surface of \OmiLup than the Fe surface abundance inhomogeneities.  Because of the simple variation of $\langle v^0 \rangle$, we argue that we only observe one surface abundance inhomogeneity close to the magnetic equator.

For the two remaining absorption lines (i.e., \ion{C}{II} and \ion{Mg}{II} lines), the measured $\langle v^0 \rangle$ followed a profile in between that of the \ion{Fe}{II} line and the \ion{Si}{II} line, albeit with a smaller amplitude.  The measured abundances of these lines are most likely following the changes in the local atmosphere caused by the Si and Fe surface abundance inhomogeneities.

\section{Discussion}
\label{sec:discussion}
Comparison between the ESPaDOnS spectroscopy and synthetic spectra did not indicate \OmiLup was an SB2 system, despite the interferometric results \citep{2013MNRAS.436.1694R}.  Either the secondary component is not visible in the spectroscopy, or the RV shifts were too small to detect due to two components of similar spectral type.  Therefore, we continue this section under the assumption that only the primary component contributes to the variability and the large-scale magnetic field.  We do comment, where applicable, what the implication would be in case of an indistinguishable secondary component in the spectropolarimetric or photometric data.

\subsection{Geometry of the magnetic field}
\label{sec:discussion_magnetic}
The rotational modulation of the measured longitudinal magnetic field favored the model a dipolar magnetic field with a quadrupolar contribution.  The strength of the quadrupolar term in the model varied with the employed LSD line mask, but it was typically about 10\,\% of the strength of the dipolar term (see Table\,\ref{tab:LSD_generalinfo}).

Assuming a typical stellar radius of 3 -- 4\,$R_{\odot}$ for a B5IV star with $T_{\mathrm{eff}}=15000$\,K \citep[e.g., Fig.\,1 of][]{2017A+A...598A..74P} and employing the measured rotation period of $2.95333(2)$\,\d and the literature $v\sin i = 27 \pm 3$\,\kms \citep{2005ESASP.560..571G}, we estimated the inclination angle to be $27\pm10\,^{\circ}$ (corresponding to an equatorial velocity of $v_{\mathrm{eq}} = 60 \pm 9$\,\kms).  These assumptions lead to an obliquity angle $\beta = 74^{+7}_{-9}\,^{\circ}$, following the scaling relation of \citet{1987AJ.....94..731S} for a purely dipolar large-scale magnetic field. The quadrupolar contribution needed for the description of the modulated longitudinal magnetic field will only have a minor effect on this estimated value.  Moreover, this value does not depend strongly on the LSD line mask used for the measurements of the longitudinal magnetic field.

A conservative lower limit for the polar strength of the large-scale magnetic field (when the geometry is a pure dipole) is 3.5 times the maximal measured longitudinal magnetic field value \citep{1967ApJ...150..547P}.  Using the measurements of the LSD profiles from the Balmer lines, we obtained a lower limit of 5.25\,kG for the polar strength of the magnetic field, since these were not influenced by the surface distribution of the chemical elements.  This value is typical for magnetic Bp stars. 

Detailed modelling of the Zeeman signatures or successful ZDI mapping of the stellar surface is required to verify the values derived from simple assumptions.  However, the current spectropolarimetric dataset is not sufficient to do so, because of missing observations at several key rotation phases.

In case the secondary component significantly contributes to the spectropolarimetric data, which was not confirmed at present, we underestimated the strength of the detected large-scale magnetic field, since the total Stokes\,I profile was used for the normalization of the $B_l$ values.  For a 40\,\% contribution of a secondary, the actual strength of the magnetic field will be underestimated by 40\,\% or 60\,\%, depending on which component is hosting the large-scale magnetic field.

We consider it unlikely that both components of the $o$\,Lup binary system host a large-scale magnetic field with the published light ratio and the similar values for $v\sin i$.  In case both hosted a large-scale magnetic field, the similar light contributions would lead to two overlapping Zeeman signatures in the LSD Stokes\,V profiles, resembling a highly complex magnetic field geometry.  The similar values for the $v\sin i$ (and light ratio) would suggest a similar rotation period, causing the longitudinal magnetic field to vary with two superimposed periods.  This would lead to a strong beating effect.  None of the above was observed (see Figs.\,\ref{fig:LSD_overlay} and \ref{fig:longmag_rotation}).  Only if both stars would rotate with exactly the same period could they pollute the observations unnoticed, and perhaps result in the necessity of the high-order fit to the measured $B_l$ values.  However, the synchronization time scale of a binary system with an orbital period of at least 20\,years is sufficiently long to exclude this possibility.

\subsection{Discrepancies in the measured magnetic field strength}
\label{sec:discussion_strength}

Strong differences in the strength of the measured longitudinal magnetic field from different chemical elements are not often noted for magnetic early-type stars.  They are, however, observed here.  Surface abundance inhomogeneities are likely the cause of these differences.  We support this hypothesis with the analysis of the LPVs, since the chemical species having a stronger longitudinal magnetic field are also the same species that indicated the rotation frequency as the dominant variability for the first moment (i.e., \ion{Fe}{II}).  Moreover, the zeroth moment of some studied absorption lines suggested that the measured equivalent width changes with the rotation phase, implying a non-uniform surface distribution for these chemical elements.  A similar conclusion was obtained by \citet{2015MNRAS.447.1418Y} for the magnetic helium-strong B2V star HD\,184927.  The larger measured $B_l$ values for LSD profiles with just \ion{Fe}{II} lines would indicate that their surface abundance inhomogeneities are located close to the magnetic pole, while the weaker fields retrieved from the \ion{He}{I} would locate their respective surface abundance inhomogeneities close to the magnetic equator.  In addition, the weaker longitudinal magnetic field measurements for the \ion{He}{I} lines could also be related to a contribution in the LSD Stokes\,I profile coming from the secondary of the \OmiLup system, which is suggested to have a similar spectral type.

We do not anticipate that the stellar pulsation is causing these severe differences in the measured longitudinal magnetic field strength. It would rather produce additional systematic offsets out of phase with the rotation period, as the pulsation frequency is not a harmonic of the rotation frequency. For the secondary component to cause these differences, it must have a different spectral type, leading to different contributions to spectral lines of different chemical species.  This is in contradiction to the binary fitting process, that suggested a similar spectral type for both components.  Also, the interferometric results indicated a relatively similar spectral type for the secondary and a mass ratio of 0.91 \citep{2013MNRAS.436.1694R}.

\subsection{Magnetosphere}
\label{sec:discussion_magnetosphere}
The detected large-scale magnetic field for the primary component of \OmiLup is sufficiently strong to create a magnetosphere.  However, the star is not sufficiently massive to have a considerable mass-loss rate, producing only a limited amount of wind material to fill the magnetosphere.  Therefore, no observational evidence of the magnetosphere is anticipated.  Indeed, no rotational modulation, nor emission features were noted for the Balmer lines.  No X-ray observations are available for \OmiLup to diagnose the interactions between wind material coming from both magnetic hemispheres.

\citet{2013MNRAS.429..398P} determined the star may host a centrifugal magnetopshere, using the magnetic properties derived by \citet{2011A+A...536L...6A}.  Repeating these computations for a $4.7\,M_{\odot}$ star with a $3.5\,R_{\odot}$ radius, the updated $P_{\mathrm{rot}} = 2.9533$\,d, and a polar magnetic field strength of $5.25$\,kG (and assuming a mass-loss rate described by \citet{2001A+A...369..574V}), we obtained $R_{\mathrm{K}} = 4.2\,R_{\star}$ and $R_{\mathrm{A}} = 26.5\,R_{\star}$ (and a mass-loss rate $\log \dot{M} = -10.42$\,dex with $\dot{M}$ given in $M_{\odot}\,\mathrm{yr}^{-1}$).  This confirms the results of \citet{2013MNRAS.429..398P} that the magnetic component of \OmiLup hosts a centrifugal magnetosphere.  Yet, as previously indicated, the mass-loss rate is too low, particularly compared to the extend of the magnetosphere ($R_{\mathrm{A}} \gg R_{\mathrm{K}}$), to produce observational evidence of magnetospheric material. In addition, the binary orbit of \OmiLup is too wide to cause effects in the circumstellar material of the magnetic component.

\subsection{Stellar pulsations}
\label{sec:discussion_pulsation}
The BRITE photometry indicated that two frequencies, namely $f_3 = 1.1057$\,\d and $f_6 = 1.2985$\,\d, were not explained as a frequency harmonic of the rotation frequency or as instrumental variability due to the spacecraft.  Furthermore, we recovered $f_3$ as the dominant periodicity in the first moment of the \ion{C}{II}\,4267.3\,$\AA$, \ion{Mg}{II}\,4481.1\,$\AA$, \ion{He}{I}\,4713.2\,$\AA$, and the \ion{He}{I}\,6678.2\,$\AA$ lines, as well as from the core of H$\alpha$.  The frequency value and the stellar parameters of the primary suggest that this frequency is a g mode.  The majority of stars exhibiting such pulsation modes show a rich frequency spectrum, with sectoral dipole modes that are quasi-constantly spaced in the period domain \citep[e.g.,][]{2014A+A...570A...8P, 2017A+A...598A..74P, 2017A+A...603A..13K}.  We only find two pulsation mode frequencies due to the limited BRITE and ground-based data sets compared to the \textit{Kepler\/} capacity in terms of aliasing.

The standard deviation of the lines (see Fig.\,\ref{fig:LPV_singlelines}) can serve as a proxy for the amplitude distribution from the pixel-by-pixel method \citep[e.g.,][]{1988ApJ...326..813G, 1997A+A...317..723T, 2006A+A...455..227Z} in case one dominant periodicity causes the LPVs.  As such, the shape of these distributions for the absorption lines that were dominantly variable with $f_3$ suggested a low-degree mode (likely a dipole mode).  Yet, detailed mode identification with \textsc{famias} did not produce conclusive results on the mode geometry.  Furthermore, the frequency $f_6 = 1.2985$\,\d was also within the appropriate frequency domain for g-mode pulsations.  In the absence of the secondary in the spectroscopy, we assumed that the g-mode pulsations originate from the magnetic component, as the majority of the LPVs were explained by $f_3$.  However, if the secondary contributed to the total flux, the periodic variability with $f_3$ or $f_6$ could be produced by the companion since its anticipated similar spectral type would place it within the SPB instability strip as well.

Without at least several detected pulsation modes, the magnetic and pulsating component of \OmiLup is not a suitable candidate for magneto-asteroseismology, and so does not provide the opportunity to investigate the influence of the large-scale magnetic field on the structure and evolution of the stellar interior.

\section{Summary and conclusions}
\label{sec:conclusions}
We combined HARPSpol and ESPaDOnS spectropolarimetry to study and characterize the large-scale magnetic field of \OmiLup.  Using the variability of the measured longitudinal magnetic field, we determined the rotation period to be $P_{\mathrm{rot}} = 2.95333(2)$\,d, which agrees with earlier estimates that the rotation period would be of the order of a few days \citep{2011A+A...536L...6A}.  We assumed that the primary component of the \OmiLup system hosts the large-scale magnetic field, given the lack of firm detection of a secondary component in the spectroscopy.

Comparing the strength of the measured $B_l$ for various chemical elements, we noted large differences, indicative of chemical peculiarity and abundance structures at the stellar surface.  The largest values were obtained for Fe, while the smallest values were derived from \ion{He}{I} lines.  This suggests that Fe surface abundance inhomogeneities are located closer to the magnetic poles, while those for He are present near the magnetic equator.  Yet, we cannot fully exclude a possible contamination by the secondary component of \OmiLup in the LSD Stokes\,I profiles.   ZDI is needed to verify the locations of the suggested surface abundance inhomogeneities.  Yet, this is not feasible with the current spectropolarimetric dataset, as we are lacking observations at several necessary rotational phases.

Fitting models to the rotational variability of the measured $B_l$ values favors a description of a dipolar magnetic field with a quadrupolar contribution.  This remains valid for the LSD profiles constructed with all metal lines, averaging out the effects of the surface abundance inhomogeneities, as well as for the LSD profiles from the Balmer lines.  Typically, the strength of the quadrupolar contribution is about 10\,\% of that of the dipolar contribution.  Using simple approximations, we estimated the inclination angle of the magnetic component of \OmiLup to be $i=27\pm10\,^{\circ}$, which then leads to an obliquity angle $\beta = 74^{+7}_{-9}\,^{\circ}$.  A conservative lower limit on the polar strength of the large-scale magnetic field, measured from the LSD profiles of the Balmer lines, would be $5.25$\,kG.

The BRITE photometry for \OmiLup shows up to six significant frequencies, indicating periodic photometric variability.  Three of these frequencies ($f_1$, $f_2$, and $f_3$) correspond to the rotation frequency, and its second and third frequency harmonic.  One frequency ($f_4$) is confirmed to be of instrumental origin, due to periodic variability of the satellite on-board temperature that was not perfectly accounted for during the correction process.  The remaining two frequencies ($f_3$ and $f_6$) fall in the frequency domain of SPB pulsations.  In case $f_3$ and $f_6$ originate from the magnetic component, \OmiLup\,A would be classified as a magnetic pulsating early-type star.  However, the few detected pulsation mode frequencies are not sufficient for detailed magneto-asteroseismic modelling.

Investigating selected absorption lines in the individual sub-exposures of the spectropolarimetric sequences indicates the presence of LPVs.  The first moment of these absorption lines almost always indicate $f_3$ as the dominant frequency, except for the \ion{Fe}{II} line where $f_{\mathrm{rot}}$ was the dominant frequency.  This is, again, suggestive of surface abundance inhomogeneities for Fe.  Moreover, the equivalent width of the studied \ion{Fe}{II} and \ion{Si}{II} lines did change significantly with the rotation phase, demonstrating non-uniform surface abundances for these chemical species.  The shape of the LPVs for the other selected absorption lines, where $f_3$ was dominant, agreed with a low-order pulsation mode, confirming that $f_3$ is a pulsation mode frequency.

\begin{acknowledgements}
B.B. thanks the participants of the third BRITE science workshop and the third BRITE spectropolarimetric workshop for the constructive comments on the presented work.  In particular, Oleg Kochukhov for his suggestion to analyse the zeroth moment in more detail.

This work has made use of the VALD database, operated at Uppsala University, the Institute of Astronomy RAS in Moscow, and the University of Vienna.

This research has made use of the SIMBAD database operated at CDS, Strasbourg (France), and of NASA's Astrophysics Data System (ADS). 

Some of the data presented in this paper were obtained from the Mikulski Archive for Space Telescopes (MAST). STScI is operated by the Association of Universities for Research in Astronomy, Inc., under NASA contract NAS5-26555. Support for MAST for non-HST data is provided by the NASA Office of Space Science via grant NNX09AF08G and by other grants and contracts.

A.\,T. acknowledges the support of the Fonds Wetenschappelijk Onderzoek - Vlaanderen (FWO) under the grant agreement G0H5416N (ERC Opvangproject).

The research leading to these results has (partially) received funding from the European Research Council (ERC) under the European Union's Horizon 2020 research and innovation programme (grant agreement N$^\circ$670519: MAMSIE) and from the Belgian Science Policy Office (Belspo) under ESA/PRODEX grant "PLATO mission development".

\end{acknowledgements}
\bibliographystyle{aa}
\bibliography{PhD_ADS}

\begin{appendix}
\section{Additional tables}
\begin{table*}
\caption{Overview of the measured longitudinal magnetic field values.}
\centering
\tabcolsep=6pt
\begin{tabular}{llcccccc}
\hline
\hline
HJD [d]	& $\phi_{\rm rot}$	& $B_l$ [G] & $B_l$ [G]		& $B_l$ [G]	&	$B_l$ [G]	& $B_l$ [G]		& $B_l$ [G]		\\
-2450000	&					& complete	& He excluded	& Balmer		& \ion{Fe}{II}	& \ion{Si}{II}	& \ion{He}{I}	\\
\hline
5704.72965	& 0.754961 &$ 400 \pm 26 $&$ 976 \pm 41 $&$ 618 \pm 142 $&$ 2038    \pm 111 $&$ 752 \pm 79 $&$ 66 \pm 42 $\\
5708.75948	& 0.119465 &$ -199    \pm 26 $&$ -478    \pm 43 $&$ -262    \pm 138 $&$ -1179   \pm 128 $&$ -140    \pm 102 $&$ -47 \pm 41 $\\
5709.73559	& 0.449976 &$ 940 \pm 27 $&$ 2243    \pm 45 $&$ 1239    \pm 143 $&$ 4770    \pm 133 $&$ 1904    \pm 83 $&$ 132 \pm 43 $\\
5709.77216	& 0.462360 &$ 908 \pm 32 $&$ 2213    \pm 51 $&$ 1177    \pm 165 $&$ 4607    \pm 147 $&$ 1707    \pm 97 $&$ 66 \pm 53 $\\
6123.56357	& 0.572476 &$ 840 \pm 22 $&$ 1976    \pm 35 $&$ 1446    \pm 127 $&$ 4285    \pm 106 $&$ 1617    \pm 65 $&$ 103 \pm 39 $\\
6124.70439	& 0.958758 &$ -133    \pm 26 $&$ -270    \pm 43 $&$ -408    \pm 138 $&$ -674    \pm 129 $&$ -167    \pm 103 $&$ -53 \pm 42 $\\
6125.45930	& 0.214371 &$ 99 \pm 29 $&$ 204 \pm 50 $&$ 513 \pm 155 $&$ 56 \pm 158 $&$ 385 \pm 105 $&$ -13 \pm 44 $\\
6125.57061	& 0.252060 &$ 206 \pm 32 $&$ 524 \pm 56 $&$ 570 \pm 170 $&$ 1167    \pm 173 $&$ 592 \pm 115 $&$ 1  \pm 48 $\\
6126.56054	& 0.587252 &$ 837 \pm 22 $&$ 1967    \pm 36 $&$ 1327    \pm 125 $&$ 4260    \pm 108 $&$ 1649    \pm 68 $&$ 118 \pm 37 $\\
6127.46812	& 0.894560 &$ 20 \pm 23 $&$ 93 \pm 38 $&$ -201    \pm 123 $&$ 268 \pm 113 $&$ -74 \pm 81 $&$ -29 \pm 37 $\\
6129.59043	& 0.613176 &$ 810 \pm 31 $&$ 1896    \pm 47 $&$ 1227    \pm 171 $&$ 3881    \pm 133 $&$ 1567    \pm 86 $&$ 92 \pm 53 $\\
6130.51806	& 0.927273 &$ -34 \pm 22 $&$ -91 \pm 34 $&$ -329    \pm 114 $&$ -206    \pm 100 $&$ -104    \pm 76 $&$ 3  \pm 36 $\\
6758.06034	& 0.413611 &$ 844 \pm 31 $&$ 1986    \pm 51 $&$ 1165    \pm 137 $&$ 4354    \pm 156 $&$ 1571    \pm 85 $&$ 106 \pm 50 $\\
6819.85180	& 0.336251 &$ 559 \pm 23 $&$ 1443    \pm 40 $&$ 967 \pm 98 $&$ 3230    \pm 130 $&$ 1158    \pm 67 $&$ -2 \pm 36 $\\
7481.61063	& 0.408343 &$ 834 \pm 26 $&$ 1948    \pm 42 $&$ 1275    \pm 139 $&$ 4148    \pm 125 $&$ 1642    \pm 80 $&$ 101 \pm 43 $\\
7481.62175	& 0.412107 &$ 820 \pm 25 $&$ 1981    \pm 41 $&$ 1343    \pm 136 $&$ 4242    \pm 122 $&$ 1658    \pm 77 $&$ 53 \pm 42 $\\
7481.63287	& 0.415871 &$ 835 \pm 25 $&$ 2018    \pm 42 $&$ 1327    \pm 137 $&$ 4391    \pm 124 $&$ 1609    \pm 78 $&$ 59 \pm 42 $\\
7481.64398	& 0.419634 &$ 845 \pm 25 $&$ 2011    \pm 41 $&$ 1249    \pm 133 $&$ 4234    \pm 119 $&$ 1621    \pm 76 $&$ 65 \pm 42 $\\
7481.65509	& 0.423396 &$ 876 \pm 26 $&$ 2043    \pm 43 $&$ 1399    \pm 141 $&$ 4228    \pm 123 $&$ 1684    \pm 81 $&$ 132 \pm 44 $\\
7481.66620	& 0.427159 &$ 878 \pm 27 $&$ 2102    \pm 43 $&$ 1252    \pm 145 $&$ 4276    \pm 124 $&$ 1733    \pm 81 $&$ 78 \pm 45 $\\
7481.67731	& 0.430921 &$ 860 \pm 28 $&$ 2109    \pm 46 $&$ 1225    \pm 154 $&$ 4440    \pm 134 $&$ 1716    \pm 86 $&$ 62 \pm 47 $\\
7481.68842	& 0.434684 &$ 856 \pm 34 $&$ 2037    \pm 54 $&$ 1393    \pm 183 $&$ 4330    \pm 157 $&$ 1604    \pm 101 $&$ 112 \pm 56 $\\
7481.82751	& 0.481777 &$ 952 \pm 47 $&$ 2201    \pm 71 $&$ 1238    \pm 262 $&$ 4835    \pm 200 $&$ 1805    \pm 136 $&$ 112 \pm 82 $\\
7481.83862	& 0.485539 &$ 861 \pm 72 $&$ 2088    \pm 109 $&$ 1526    \pm 396 $&$ 5059    \pm 335 $&$ 1639    \pm 209 $&$ 24 \pm 127 $\\
7482.65216	& 0.761005 &$ 434 \pm 39 $&$ 910 \pm 62 $&$ 611 \pm 219 $&$ 1989    \pm 172 $&$ 758 \pm 124 $&$ 96 \pm 66 $\\
7482.66327	& 0.764767 &$ 401 \pm 46 $&$ 862 \pm 73 $&$ 758 \pm 253 $&$ 1932    \pm 206 $&$ 777 \pm 146 $&$ 91 \pm 77 $\\
7484.69562	& 0.452922 &$ 842 \pm 37 $&$ 2083    \pm 61 $&$ 1535    \pm 207 $&$ 4765    \pm 177 $&$ 1747    \pm 121 $&$ 63 \pm 62 $\\
7484.70673	& 0.456684 &$ 815 \pm 35 $&$ 2039    \pm 58 $&$ 1257    \pm 201 $&$ 4444    \pm 167 $&$ 1652    \pm 115 $&$ 76 \pm 58 $\\
7484.79338	& 0.486023 &$ 884 \pm 28 $&$ 2145    \pm 46 $&$ 1267    \pm 146 $&$ 4486    \pm 130 $&$ 1893    \pm 91 $&$ 67 \pm 47 $\\
7484.80449	& 0.489786 &$ 884 \pm 29 $&$ 2147    \pm 47 $&$ 1309    \pm 159 $&$ 4576    \pm 133 $&$ 1842    \pm 93 $&$ 97 \pm 48 $\\
7484.88788	& 0.518020 &$ 956 \pm 27 $&$ 2179    \pm 44 $&$ 1392    \pm 156 $&$ 4677    \pm 125 $&$ 1876    \pm 86 $&$ 165 \pm 47 $\\
7484.89899	& 0.521782 &$ 935 \pm 27 $&$ 2204    \pm 43 $&$ 1481    \pm 157 $&$ 4676    \pm 123 $&$ 1830    \pm 84 $&$ 62 \pm 47 $\\
7485.73598	& 0.805188 &$ 264 \pm 20 $&$ 649 \pm 34 $&$ 164 \pm 107 $&$ 1174    \pm 92 $&$ 382 \pm 72 $&$ 30 \pm 32 $\\
7485.74709	& 0.808952 &$ 254 \pm 20 $&$ 610 \pm 33 $&$ 212 \pm 107 $&$ 1128    \pm 91 $&$ 371 \pm 71 $&$ 25 \pm 33 $\\
7485.83035	& 0.837143 &$ 175 \pm 22 $&$ 431 \pm 35 $&$ 222 \pm 113 $&$ 744 \pm 93 $&$ 194 \pm 78 $&$ 18 \pm 36 $\\
7485.84146	& 0.840905 &$ 201 \pm 23 $&$ 472 \pm 37 $&$ 126 \pm 120 $&$ 875 \pm 98 $&$ 339 \pm 83 $&$ 14 \pm 38 $\\
\hline
\hline
\end{tabular}
\label{tab:appendix_Bl_values}
\tablefoot{For each observation, the HJD at mid-exposure, the corresponding rotation phase, $\phi_{\rm rot}$, calculated using $P_{\rm rot} = 2.95333$\,d and $T_0 = \mathrm{HJD}\,2455702.5$, and the measured longitudinal magnetic field $B_l$ for various LSD line masks are indicated.}
\end{table*}

\end{appendix}
\end{document}